\documentclass[11pt]{article}
\usepackage{color}
\usepackage{times}
\usepackage[T1]{fontenc}
\usepackage[latin9]{inputenc}
\usepackage{graphicx}
\usepackage{babel}
\usepackage{amssymb}
\usepackage[a4paper, total={6in, 8in}]{geometry}

\graphicspath{ {./figs/} } 

\begin{document}

\title{
A biological hydraulic accumulator: How the squirting cucumber, \emph{Ecballium elaterium}, squirts its seeds 
}

\author{Sergio Test\'on-Mart\'{\i}nez$^{1}$, Carlos Gutierrez-Ariza$^{1}$,  Francisco J. Oca\~na$^{2}$, \\
 Rafael Rubio de Casas$^{2,3}$,  C. Ignacio Sainz-D\'{\i}az$^{1}$, Julyan H. E. Cartwright$^{1,4}$
\\$^1$Instituto Andaluz de Ciencias de la Tierra, CSIC, \\18100 Armilla, Granada, Spain \\$^2$Departamento de Ecolog\'{\i}a, \\Universidad de Granada, 18071 Granada, Spain \\$^3$Research Unit ``Modeling Nature'' MNat,  \\Universidad de Granada, 18071 Granada, Spain 
\\$^4$Instituto Carlos I de F\'{\i}sica Te\'orica y Computacional, \\Universidad de Granada, 18071 Granada, Spain}
 
\date{Version of: \today}

\maketitle

\begin{abstract}
Seed dispersal is a fundamental process that allows offspring to reach suitable habitats and colonize new environments. While most plants rely on external vectors, some have evolved mechanisms that employ the buildup of liquid pressure in a closed compartment and its explosive release to disperse their seeds. This form of energy storage, reinvented by humans for engineering applications, is termed a hydraulic accumulator. Here we investigated the fluid mechanics involved in dispersal in the squirting cucumber, \emph{Ecballium elaterium} integrating high-speed videography (up to 10\,000 fps), microtomography, and internal pressure sensors. We recorded long-term pressure time series showing that \emph{E. elaterium} exhibits circadian (24-hour) and ultradian (short-period) rhythms. Remarkably, the measurements revealed a lack of correlation between fruit and stem turgor; while the stem showed strong circadian cycles, the fruit often did not, suggesting isolated physiological processes in different tissues. The fruit's spongy wall tissue stores elastic potential energy as turgor pressure builds to nearly one atmosphere (92--99 kPa). Upon detachment, this energy is rapidly released to expel a turbulent, particle-laden liquid jet. Microtomography revealed that the seeds are packed around a central funiculus, a configuration that optimizes their exit through the basal orifice at velocities of up to 30 m/s. Seeds eventually move faster than the liquid droplets during the later stages of ejection as they shed their liquid coating. This sophisticated mechanism ensures a broad dispersal cone, effectively spreading offspring across space and environmental conditions
\end{abstract}

\section{Introduction}

For plants, as for other living organisms, the management of liquid pressure is vital to their existence. While in single-celled organisms the focus is on controlling osmotic pressure across the cell wall, in vascular plants liquid pressure also needs to be managed on an organism-wide basis in order to pump energy-transporting fluid --- sap --- from place to place \cite{Bruhn2014,Jensen2016}. Most plants move only slowly compared to animals, with their movements under so-called turgor (osmotic) pressure constrained by physics to be relatively slow \cite{Skotheim2005,Dumais2012,Forterre2013,Forterre2016}.  However, some species have evolved ways to bypass this physical limitation by utilizing other forms of energy. The example we investigate here in the squirting cucumber, {\it Ecballium elaterium}, is the buildup of  pressure in a closed compartment that leads the explosive detachment of the seed pods and the dispersal of its seeds in a liquid jet.

\subsection{Dispersal}

Dispersal is a fundamental biological process \cite{howe1982ecology,nathan2000spatial}. The fitness of any organism relies on its capacity to reach a suitable habitat for development and reproduction. By distributing progeny across varying spatial or temporal conditions, dispersal can facilitate survival across time and space \cite{bowler2005causes}. It serves as a means of colonizing new environments, contributes to population maintenance and expansion, and reduces competition among neighbouring plants by promoting escape from crowding \cite{howe1982ecology,levin2003ecology}. Furthermore, spatial dispersal is vital for species persistence, enabling them to track shifting habitat boundaries through migration, which can be critical during periods of climate change \cite{travis2013dispersal,corlett2013will}. 

In plants, dispersal is not under the control of the dispersed individual, which is often the seed; instead, characteristics of the propagule such as morphology, size, and phenology are primarily maternal, and dispersal results from the interaction between the maternally-shaped propagule and environmental factors, including dispersal vectors \cite{schupp2019intrinsic}. This interplay between propagule traits and environmental forces leads to a wide array of specific dispersal mechanisms. These diverse strategies are often categorized into dispersal \emph{syndromes} based on the vector, such as anemochory (wind dispersal) \cite{nathan2000spatial}, endozoochory (animal dispersal; often via fleshy fruits) \cite{barea2022evolution}, and hydrochory (water dispersal) \cite{soomers2013wind}. 

Among these syndromes, plants that do not rely on an external dispersal vector are considered \emph{autochorous} \cite{cousens2008dispersal}. These self-dispersing species have evolved remarkable internal mechanisms to propel their propagules, often leading to highly specialized and energetically demanding processes. One particular case of autochory is that of seeds released by explosive dehiscence of the fruit (ballochory or ballistic dispersal), commonly due to increased turgor pressure or hygroscopic tension \cite{Skotheim2005,hayashi2009mechanics}. This dispersal syndrome often results in exceptionally fast movements, particularly when osmotic processes are coupled with the rapid release of stored elastic energy. 

Explosive dehiscence, or ballochory, is observed across numerous plant families, including the Balsaminaceae, Geraniaceae, and Euphorbiaceae, representing a convergent evolutionary strategy for effective self-dispersal \cite{stamp1983ecological}. The dispersal mechanism of the squirting cucumber, \emph{Ecballium elaterium},  one of the most rapid motions in the plant kingdom, is a clear example of this phenomenon. \emph{E. elaterium} relies on internal hydrostatic pressure to explosively eject its seeds via a high-pressure liquid jet. This unusual, forceful self-dispersal mechanism allows the plant to launch its propagules over substantial distances and ensures a broad spatial distribution of offspring, serving as a crucial stabilizing force in population dynamics by effectively reducing intraspecific competition and escape from crowding \cite{howe1982ecology,levin2003ecology}. It is a 
 botanical instance of the original sense of the term \emph{broadcasting}.

\subsection{Squirting cucumber, {\it Ecballium elaterium}}

{\it Ecballium elaterium}, the squirting cucumber,  
has been known  throughout the Mediterranean region since antiquity  \cite{janick2007} for its impressive manner of jet-propelled seed dispersal.  The seed pod detaches explosively from its stalk as a jet of liquid squirts from the opening, and the seeds are expelled with the jet to land some metres away. 
There are first century CE descriptions of this process by  Pliny the Elder in his {\em Historia Naturalis} \cite{pliny} and by Dioscorides in {\em De Materia Medica} \cite{dioscorides}.
Moving now to the 20th century,  Guttenberg \cite{guttenberg} made an in-depth study of the mechanism,
Ziegenspeck \cite{ziegenspeck} investigated the relationship with the properties of the cell walls, 
and Overbeck \cite{overbeck1930druckkraften} made extensive research on the pressures involved.
Obaton \cite{Obaton1947} filmed the seed pod jet at 64 frames per second in 1947 and  experimented with the pods to estimate that a pressure of 72~kPa was necessary to produce such a jet. He hypothesized that the somewhat conical form of the orifice of the pod maximized the seed dispersal. Lewes \cite{Lewes1951} measured somewhat lower pressures of  22~kPa in unripe seed pods in 1951. He believed that his technique was underestimating, but that Obaton's technique was overestimating the osmotic pressure. Wolters \cite{wolters1963} again filmed {\it Ecballium} jets \cite{wolters_film}, but now in 1963 at 8000 frames per second, and found peak velocities of the seeds over 16~m/s ($\sim 60$~km/h).

While the remarkable mechanical efficiency and descriptive accounts of ballistic dispersal in \emph{E. elaterium} are well established, quantitative understanding of the precise biophysical parameters governing seed ejection dynamics, particularly under varying environmental conditions or developmental stages, remains limited.

\subsection{Pressure-driven seed dispersal in plants}

The seed projection mechanism of \emph{E. elaterium}, as almost all plant movements, is constrained by the velocity of water exchange by diffusion through the plant tissue, the poroelastic timescale over a tissue of size $L$ \cite{Skotheim2005,Dumais2012,Forterre2013,Forterre2016}
$\tau_p\sim L^2/D\sim\eta L^2/(k E)$,
where 
$D=kE/\eta$ is a diffusion coefficient depending on the Darcy permeability $k$ of the soft porous medium that is plant tissue,
$E$ is Young's modulus of the medium and $\eta$ is the viscosity of the fluid. In the present case, given typical values of the diffusion coefficient for plant tissues, $D\sim 10^{-9}$~m$^2$~s$^{-1}$,  for a seed pod of the order of a few centimetres, this timescale is approaching $10^5$~s; approximately a day.

A few plants and fungi, however, have coupled osmotic processes with the storage of energy in elastic material \cite{Skotheim2005,Dumais2012,Forterre2013,Forterre2016}. This elastic potential energy, slowly stored, may be rapidly released, as in a catapult, allowing them to achieve sudden movement in order to disperse seeds \cite{Sakes2016} or to catch insects \cite{Guo2015}. Some of these rapid movements rely on reversible buckling of elastic material; others on irreversible fracture.

Sakes et al.\ \cite{Sakes2016}  discuss the fungi Ascomycota and \emph{Pilobolus}, in which fluid-filled sacs eject spores. In both cases, osmotic pressure increases in a closed compartment containing spores, which eventually detaches leading to their ejection at high speeds. They also discuss plants including the genus \emph{Arceuthobium}, dwarf mistletoe, which has a similar mechanism of increasing osmotic pressure in the fruit leading to fracture between the stem and the base of the fruit, thus discharging the seeds.
Also in dwarf mistletoe, deBruyn et al.\ \cite{deBruyn2015} further found a heating of some 2K above ambient temperature just before the explosion event. They concluded that ``endogenous heat production acts as an internal trigger for the final explosive fracture of the fruit''.  
Hofhuis et al.\ \cite{Hofhuis2016} looked at \emph{Cardamine hirsuta}, hairy bittercress, another plant using an explosive mechanism for seed dispersal. They found the mechanism to be further case of the fruit exploding owing to turgidity, i.e., to osmotic pressure, but in this case involving  tension produced by differential contraction of fruit wall tissues.
A similar mechanism was shown by Westermeier et al.\ \cite{westermeier2018carnivorous} for the carnivorous waterwheel plant \emph{Aldrovanda vesiculosa}.

\begin{figure}[tbp]
\begin{center}
\includegraphics[width=0.49\columnwidth]{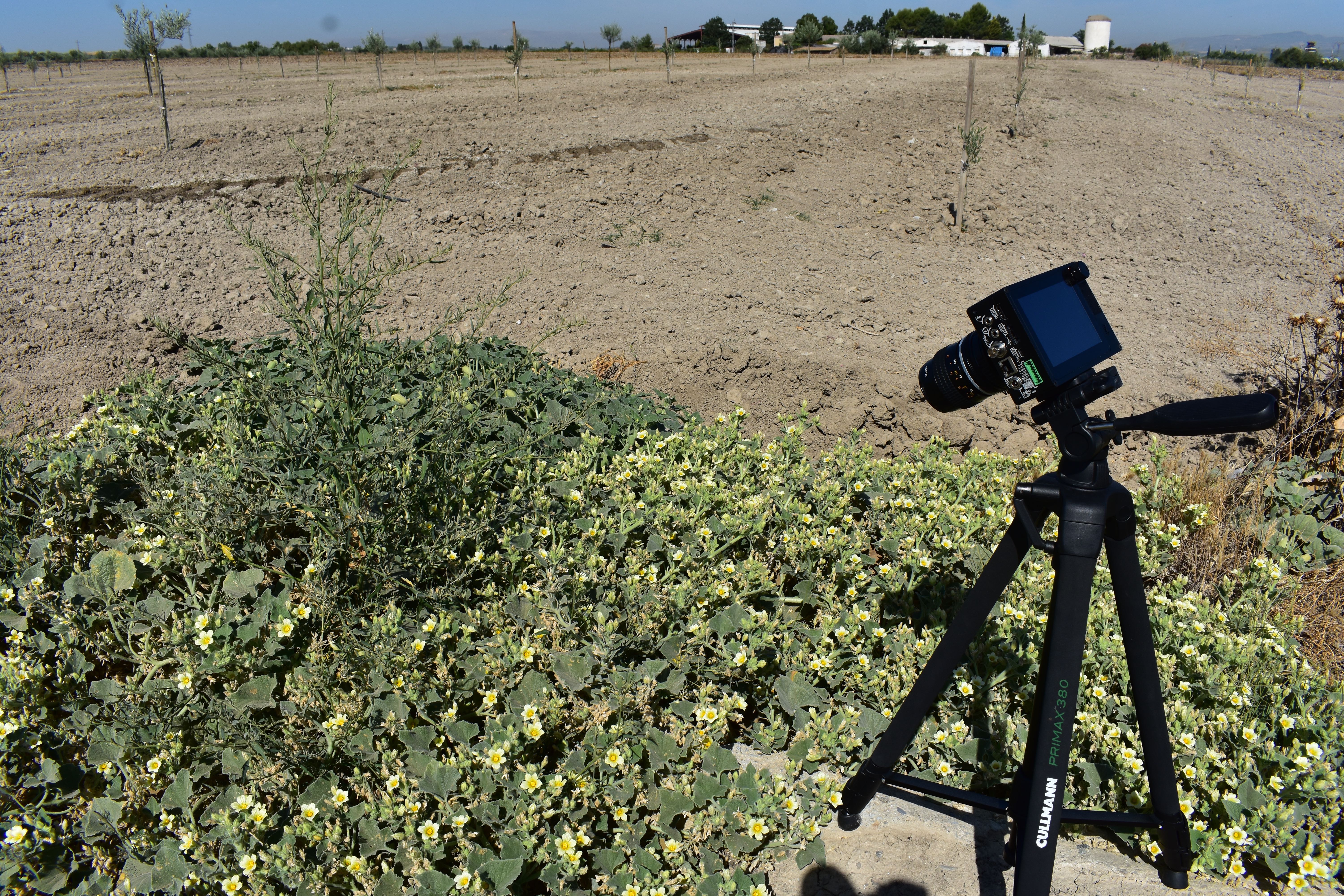}
\includegraphics[width = 0.49\columnwidth]{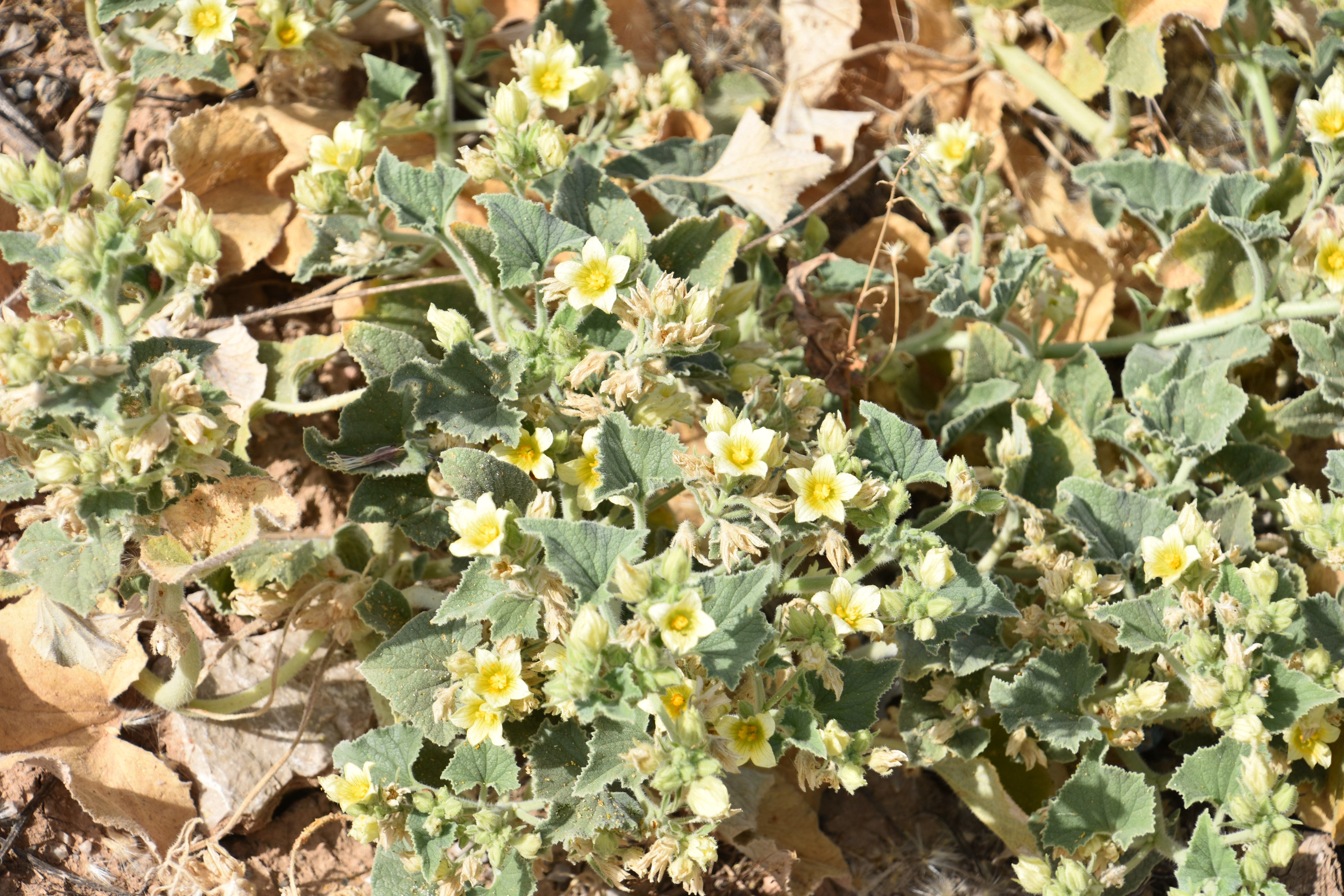}
\includegraphics[width = 0.433\columnwidth]{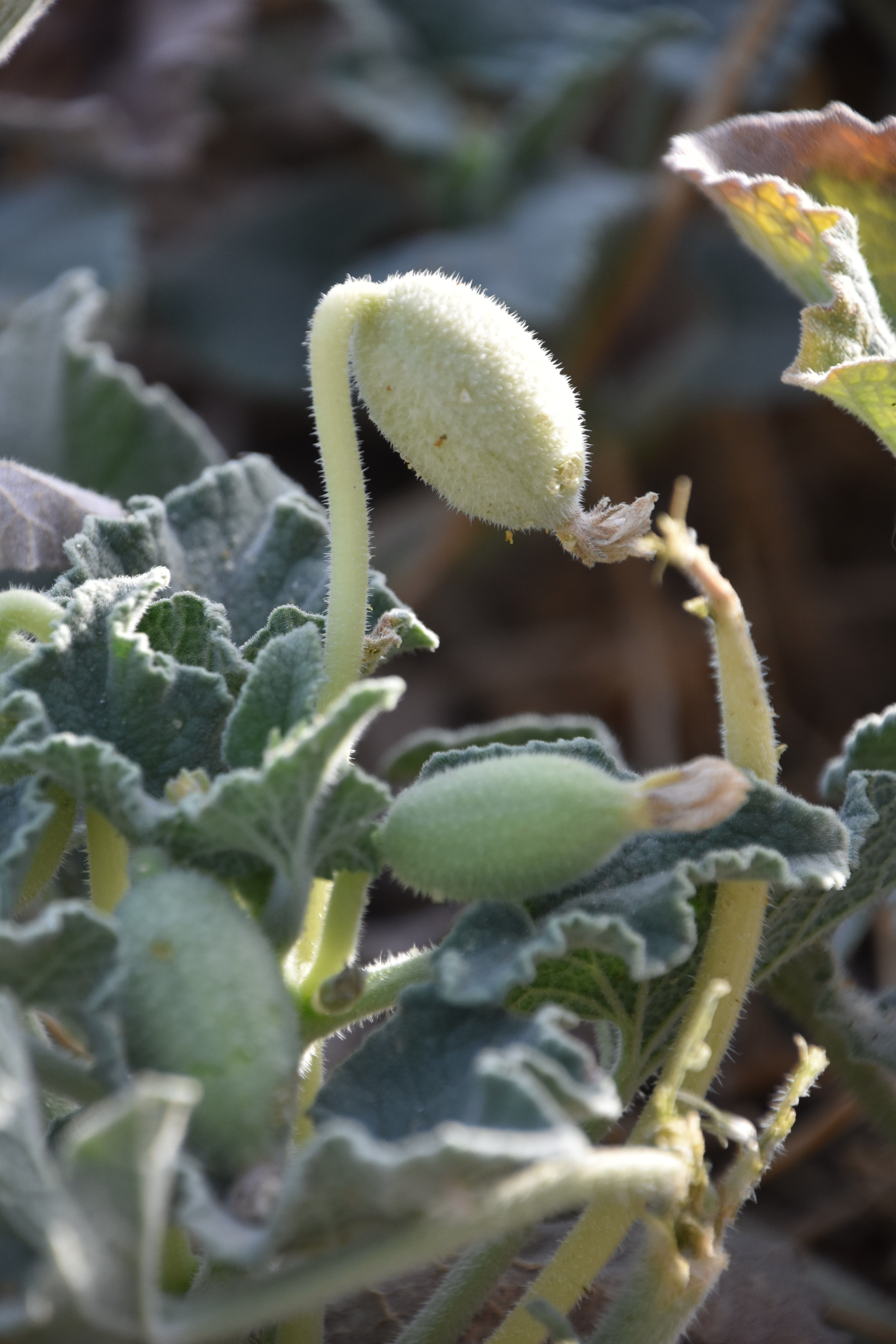}
\includegraphics[width=0.555\columnwidth]{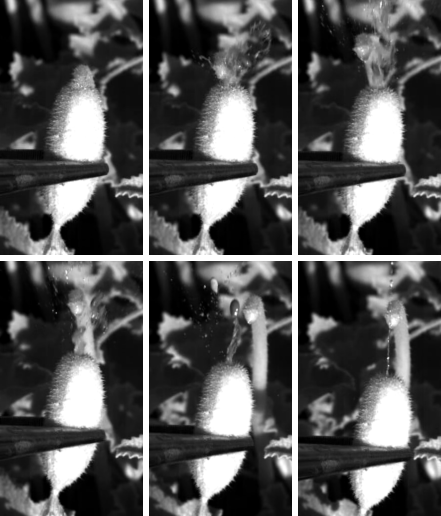}
\end{center}
\caption{{\it Ecballium elaterium}, from the large to the small scale: 
(a) field setup used to capture seed pod explosions; 
(b) squirting cucumber plants in flower;
(c) ripe   seed pods;
(d) sequence showing explosion of seed pod.
}\label{fig:field_trip}
\end{figure}

\section{Materials and methods}

To study experimentally the explosive detachment of the seed pods of   {\it E. elaterium}
we adopted two complementary approaches: On one hand pressure measurements on the fruit and
the stem to characterize the ripening process
and, on the other hand, 
the study of the fluid and seed dynamics via video recording of the explosions. Alongside these, we used a third technique, microtomography, to characterize the geometry of the seed pod.

\paragraph{Microtomography}

Microtomography was performed on intact seed pods using high resolution X-ray microscopy of  resolution 1~$\mu$m with a   Xradia 510 VERSA ZEISS sited at the Centre for Scientific Instrumentation of the University of Granada.
A Freeze dryer FLEXI-DRY-$\mu$P (FTS systems) was used for  lyophilization.

Data treatment was performed using the software kit FIJI (Fiji Is Just ImageJ), as a quick prototyping starting point and then pipelined via Python.
We  performed morphological operations on the binarized image of the fruit to remove parts to be able to examine separately the various structures.  
A combination of erosions and dilations allowed us to remove internal structure, and focus on characterizing the outer wall in order to be able to remove it later from the reconstruction. 

For seed detection and characterization we used an extension of Hough's transform to detect 3D ellipsoidal hollow shells. A template kernel is generated as an ellipsoidal hollow shell and then the 3D volume is convolved with it using Fast Fourier Transform acceleration resulting in a Hough map indicating the centres of structures that best match an ellipsoidal shell geometry.
Given the centres and their perimetric voxels we  applied a watershed segmentation algorithm to determine how far a seed extends from its centre and to separate voxels that are at the interface of two touching seeds, thus getting all seed-like structures characterized.

\paragraph{Video recording}

Two wild populations of squirting cucumber
were visited in August 2018   ---  Fig.~\ref{fig:field_trip} --- in
order to record  explosions in the field.
One location was 
 in  Deifontes (37\textdegree 19'23.0"N 3\textdegree 35'49.8"W);  the other  was in  the town of Santa Fe (37\textdegree 10'45.1"N 3\textdegree 45'34.1"W).
To capture the detachment of the seed pods a Chronos 1.4 high speed camera
(Kron Technologies Inc.) equipped with a macro lens was used. Depending
on lighting conditions and the position of the fruit in the plant (i.e., whether it was
relatively isolated from the rest of the plant or in its interior) a range of
frame-rates was used.
For situations with the most natural light we could use higher shutter speeds and went up
to 10\,000~fps whereas for more shaded situations we needed to lower the recording
speed to a minimum of 5\,000~fps; with less than that we would miss
part of the dynamics we are interested in. After setting up the camera a sheet of ruled paper was located behind a ripe pod and the pod was exploded with a very light touch with a pair of tweezers that were also to hold the pod in place for videography.
A  particle tracking algorithm
was written in Python and applied to the video films to obtain information about
seeds and pod trajectories to be able to study the relative kinematics.

\paragraph{Liquid pressure measurement}

Plants were collected in August 2018 both from the locations listed above and also from   V\'elez de Benaudalla (36\textdegree 49'53.6"N 3\textdegree 30'36.1"W). These plants were collected along with a certain portion of their long main tap root for ensuring their survival in the transplantation procedure, and were placed in portable pots of $30\times 30\times 60$ cm filled in part with the same soil of the collection place together with additional commercial potting soil. These pots were kept in a greenhouse of the Department of Ecology of the Universidad de Granada under controlled temperature ($25 \pm 5$\textdegree C) and relative humidity (approx. $60\%$).
Two pressure sensors manufactured by Pasco were used, one for
the pod (Pasco PS-3203 wireless pressure sensor) and a more sensitive one
for the stem (Pasco  PS-2114 relative pressure sensor).
An appropriate arrangement of needles and plastic tubes  allowed us
to measure liquid pressures inside the pod and stem to obtain data about the
pressure causing the burst and the changes in pressure over time
in the plant in general. 

We calculated both Fourier transforms and the Welch spectral density estimation algorithm that  reduces the variance of the spectral density estimate using
the average of periodograms of multiple overlapping segments of the signal \cite{welch1967use}.

\section{Experimental results}

\subsection{Tomography}

\begin{figure}[tbp]
\includegraphics[width=\columnwidth]{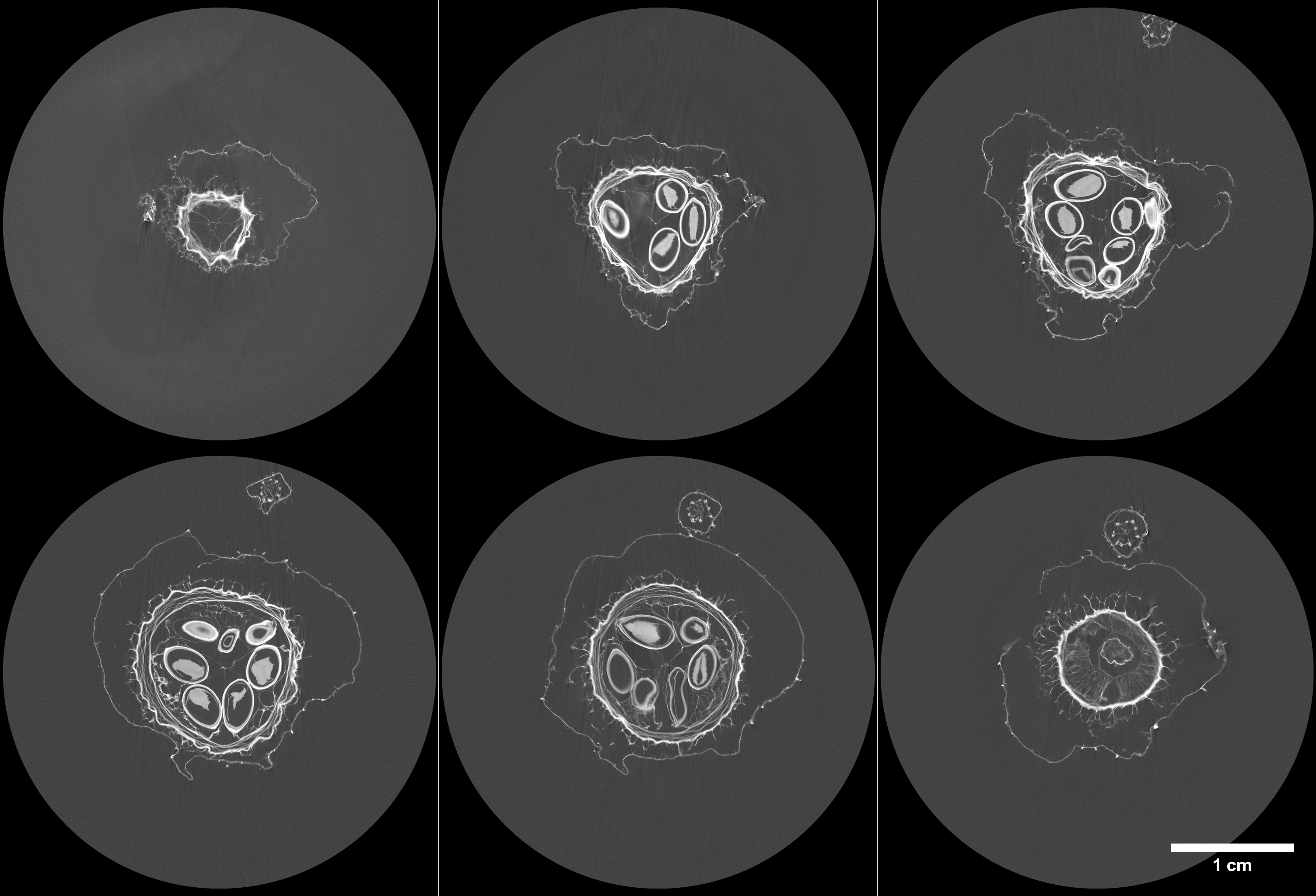}
\includegraphics[width=\columnwidth]{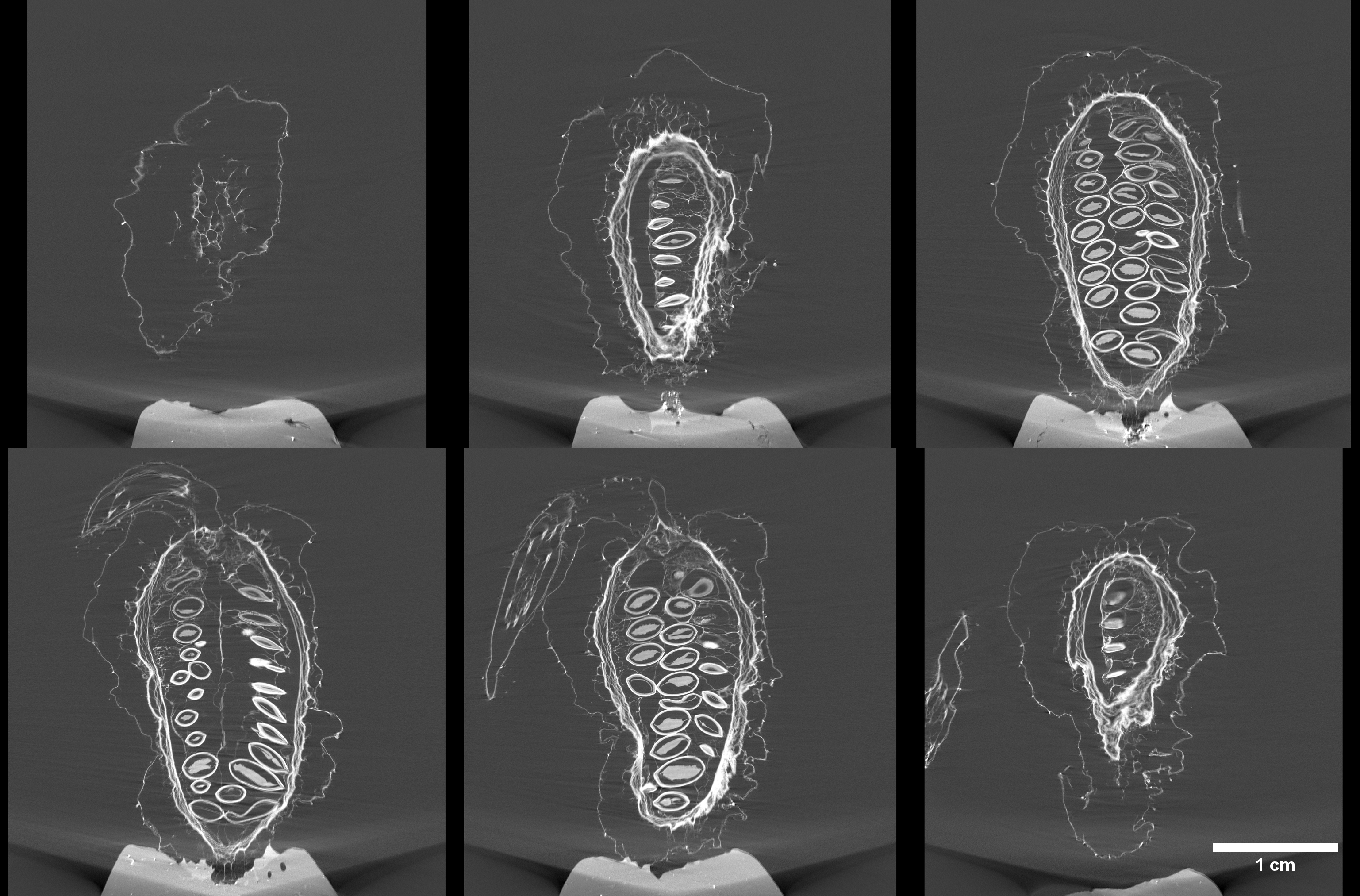}
\caption{Tomography: transverse and longitudinal  slices showing the disposition of seeds within the fruit.
}\label{fig:tomography}
\end{figure}

\begin{figure}[tbp]
\centering
\includegraphics[width=0.505\columnwidth]{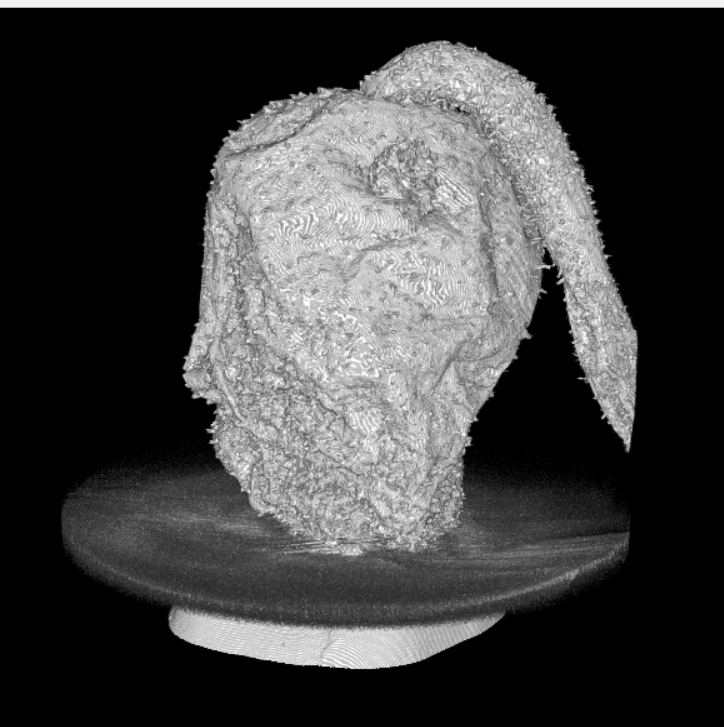}
\includegraphics[width=0.485\columnwidth]{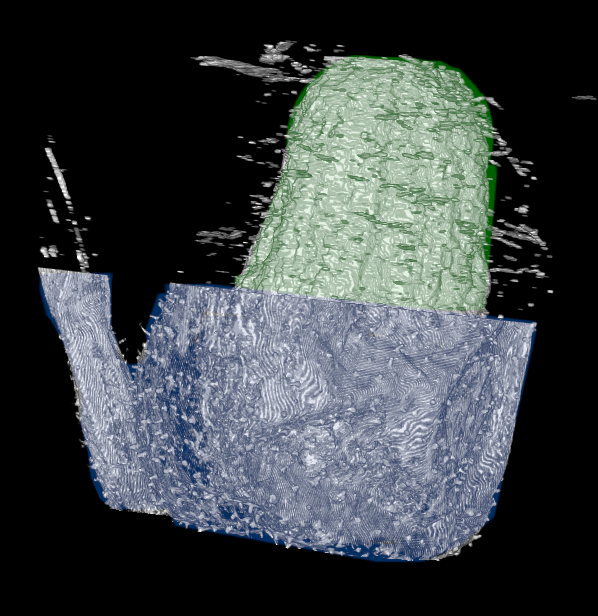} \\
\includegraphics[width=\columnwidth]{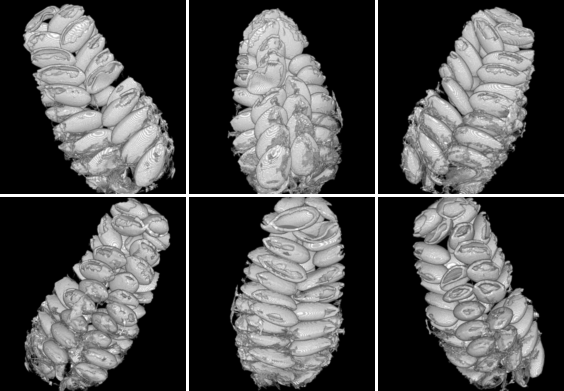}
\includegraphics[width=\columnwidth]{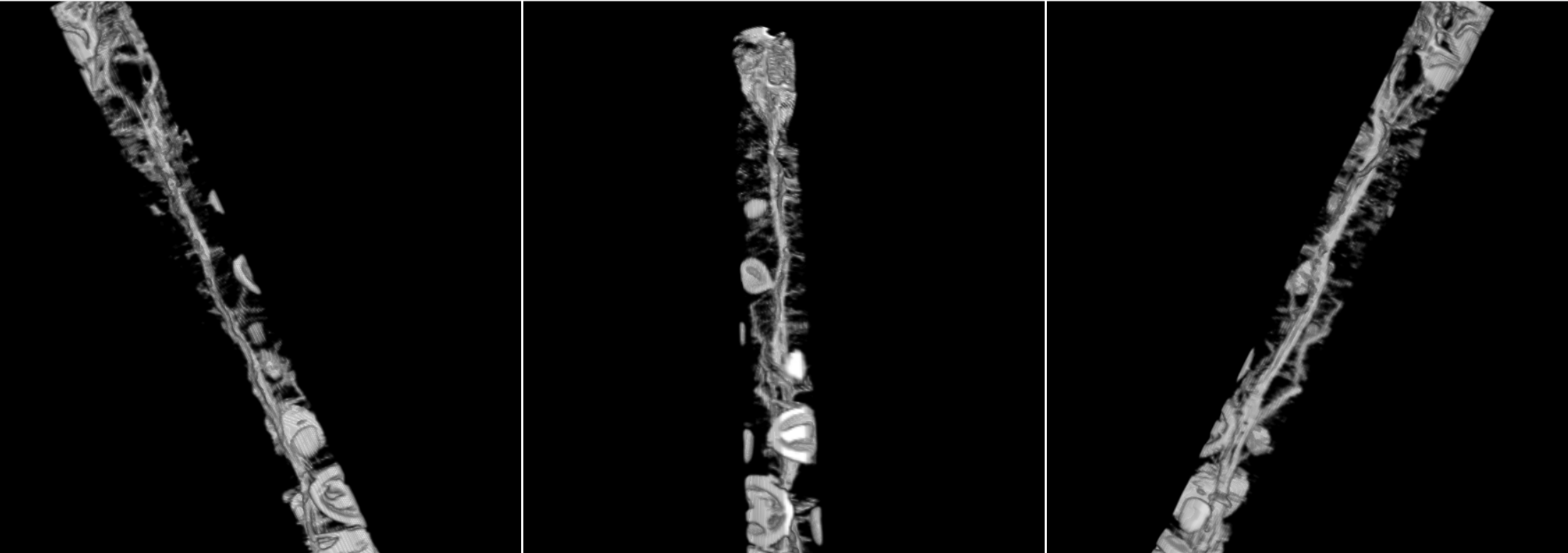}
\caption{Tomography: 3D images. 
(a) Overall aspect;
(b) Cut-away showing width of pressure vessel wall,
(c) overall organization of seeds, and 
(d) showing the central funiculus.}
\label{fig:tomography2}
\end{figure}

We took  tomographic
scans of an intact seed pod that had undergone lyophilization. From these data we can extract  slices, Fig.~\ref{fig:tomography}, and reconstruct 3D images, Fig.~\ref{fig:tomography2}.
The wall is seen to be both thick and composed of materials of differing radiopacity. 
The inner wall of the pod is stretched due to pressure.

After isolating the interior wall of the fruit with digital processing we can obtain reconstructions of the fluid chamber with and without its inner structure. 
In Fig.~\ref{fig:tomography2}(c) we  show the 3D reconstruction the internal chamber, seeds included.
The seeds are arranged in a very ordered fashion within the pod. The 3D scan also reveals the funiculus in the centre of the pod to which all the seeds attach; Fig.~\ref{fig:tomography2}(d). 

Measurement gives  an estimate for the total internal volume of the chamber of 1.21~ml, of which 0.43~ml is occupied by seeds, and 0.78~ml by liquid. 
The pod contains approximately 45 seeds, which have dimensions 5.0$\pm$2.0 \texttimes~3.0$\pm$1.1 
\texttimes~2.0$\pm$0.9~mm. 
We also weighed the seeds; the mean is some 14~mg (minimum 6~mg, maximum 18~mg). The seeds are not fully axisymmetric, and have a rounder end, at the chalaza, and a more pointed end, at the micropyle.

\subsection{Pressure measurements}

\begin{figure}[tbp]
\begin{center}
\centering\includegraphics[width = \columnwidth]{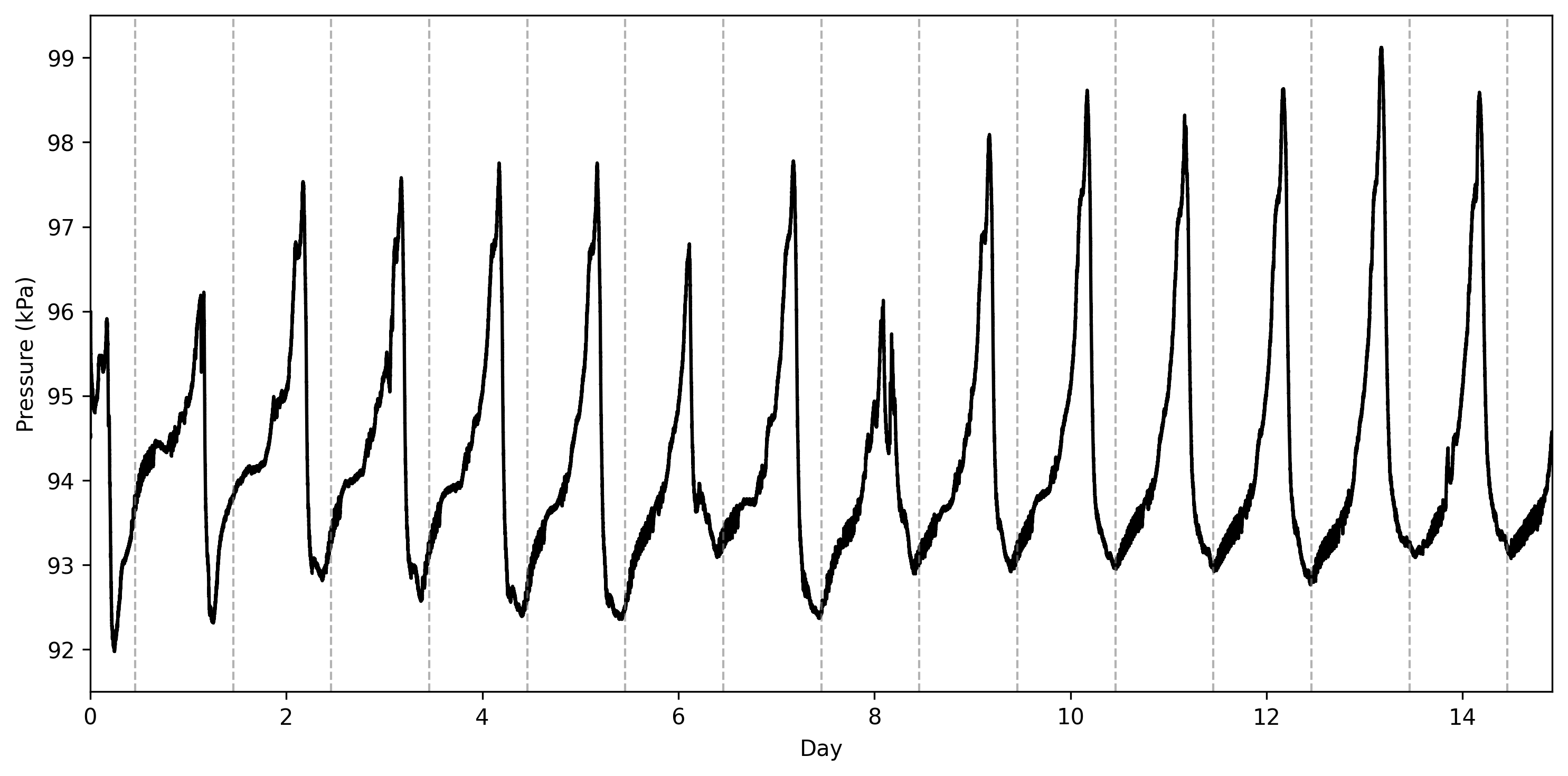}
\includegraphics[width=\columnwidth]{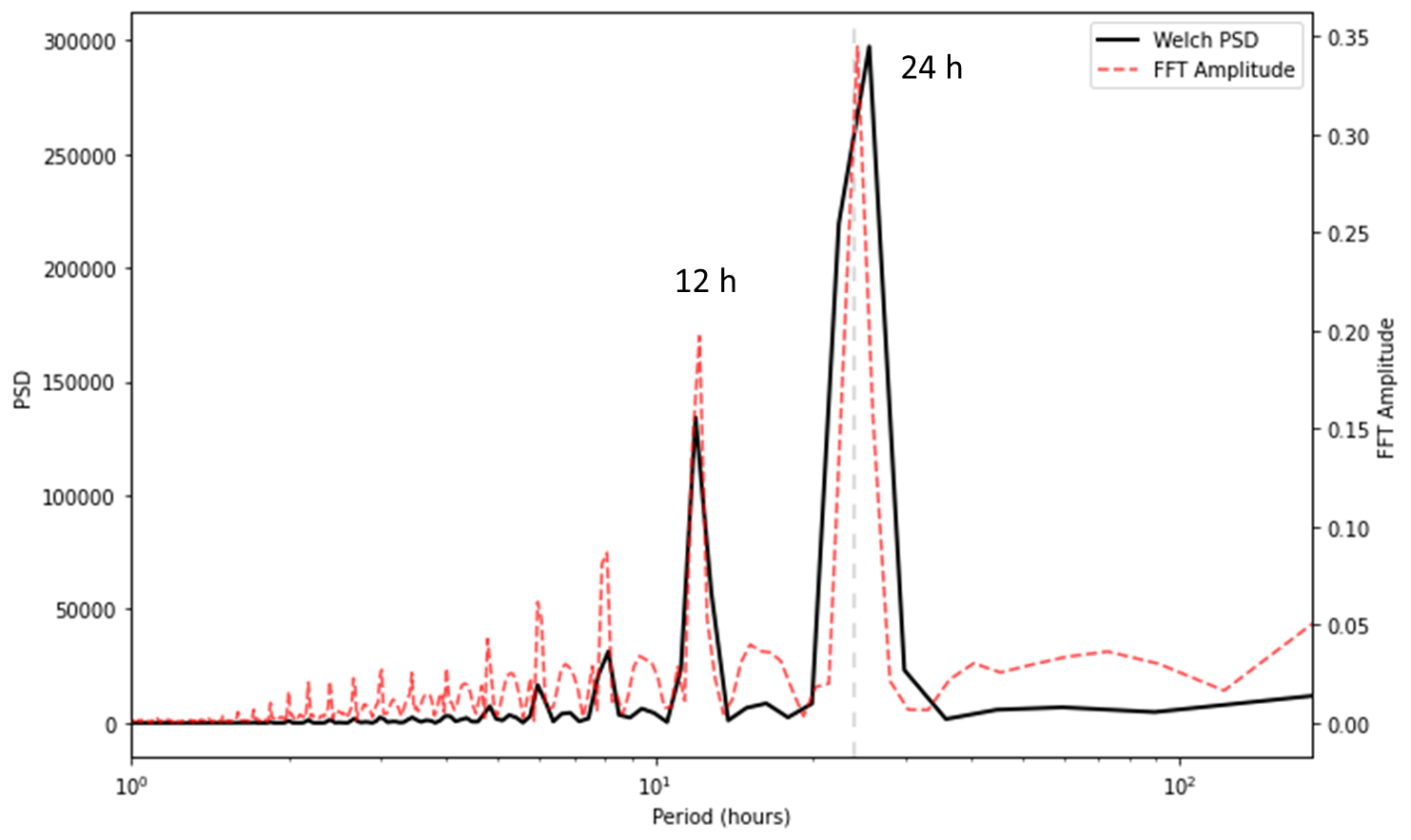}
\end{center}
Figure 3: (a)
\end{figure}

\begin{figure}[tbp]
\begin{center}
\centering\includegraphics[width = \columnwidth]{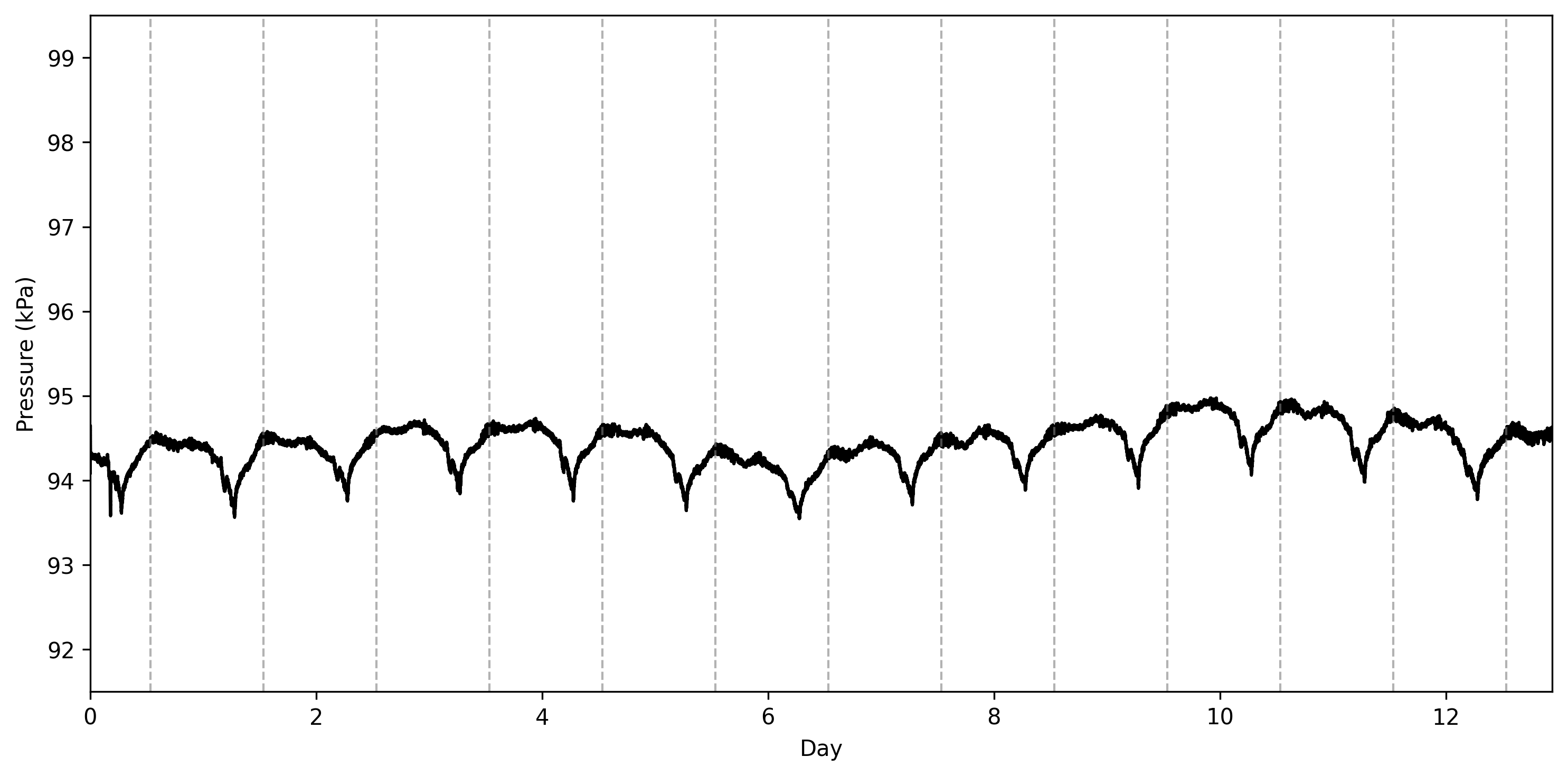}
\includegraphics[width=\columnwidth]{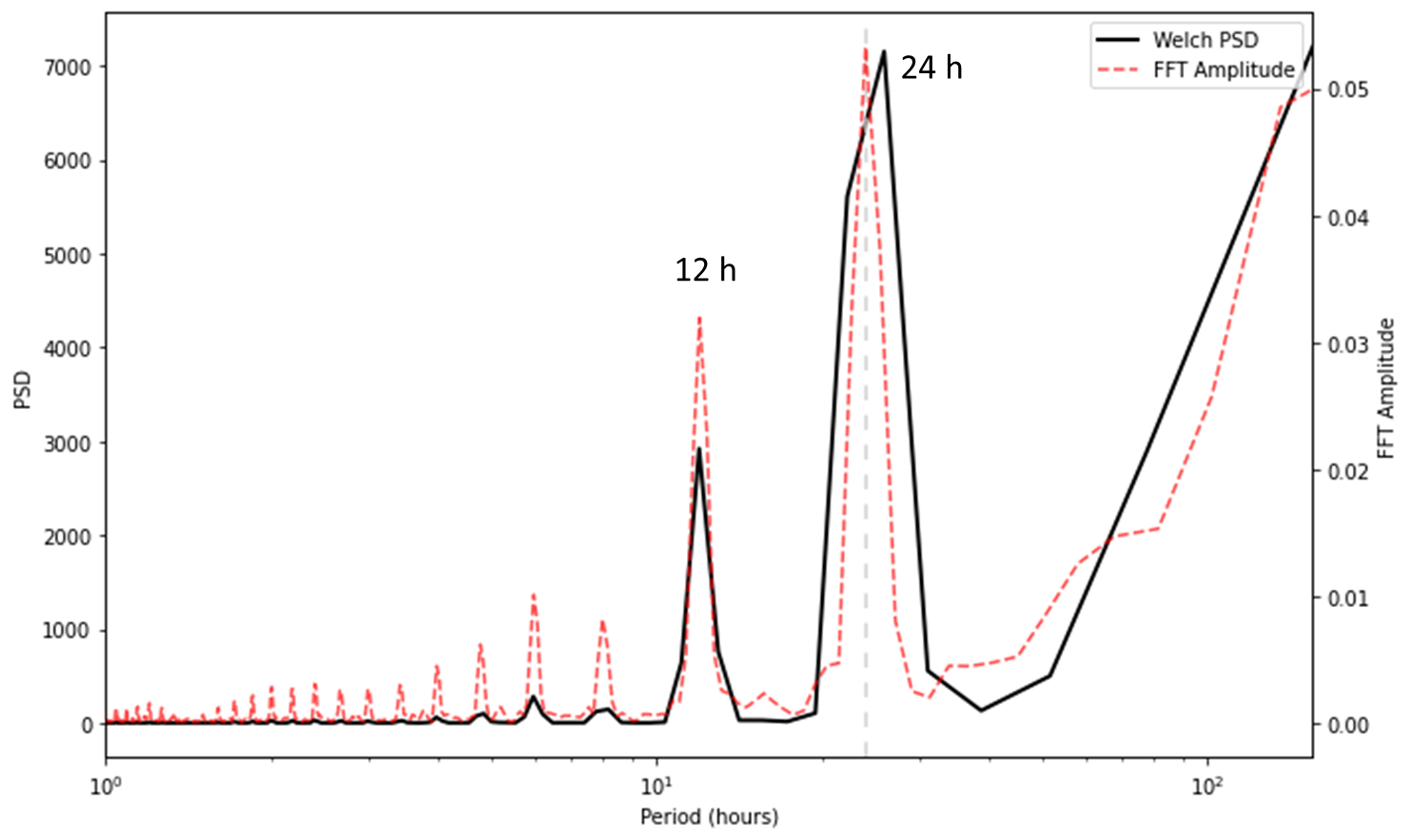}
\end{center}
Figure 3: (b) 
\end{figure}

\begin{figure}[tbp]
\begin{center}
\centering\includegraphics[width = \columnwidth]{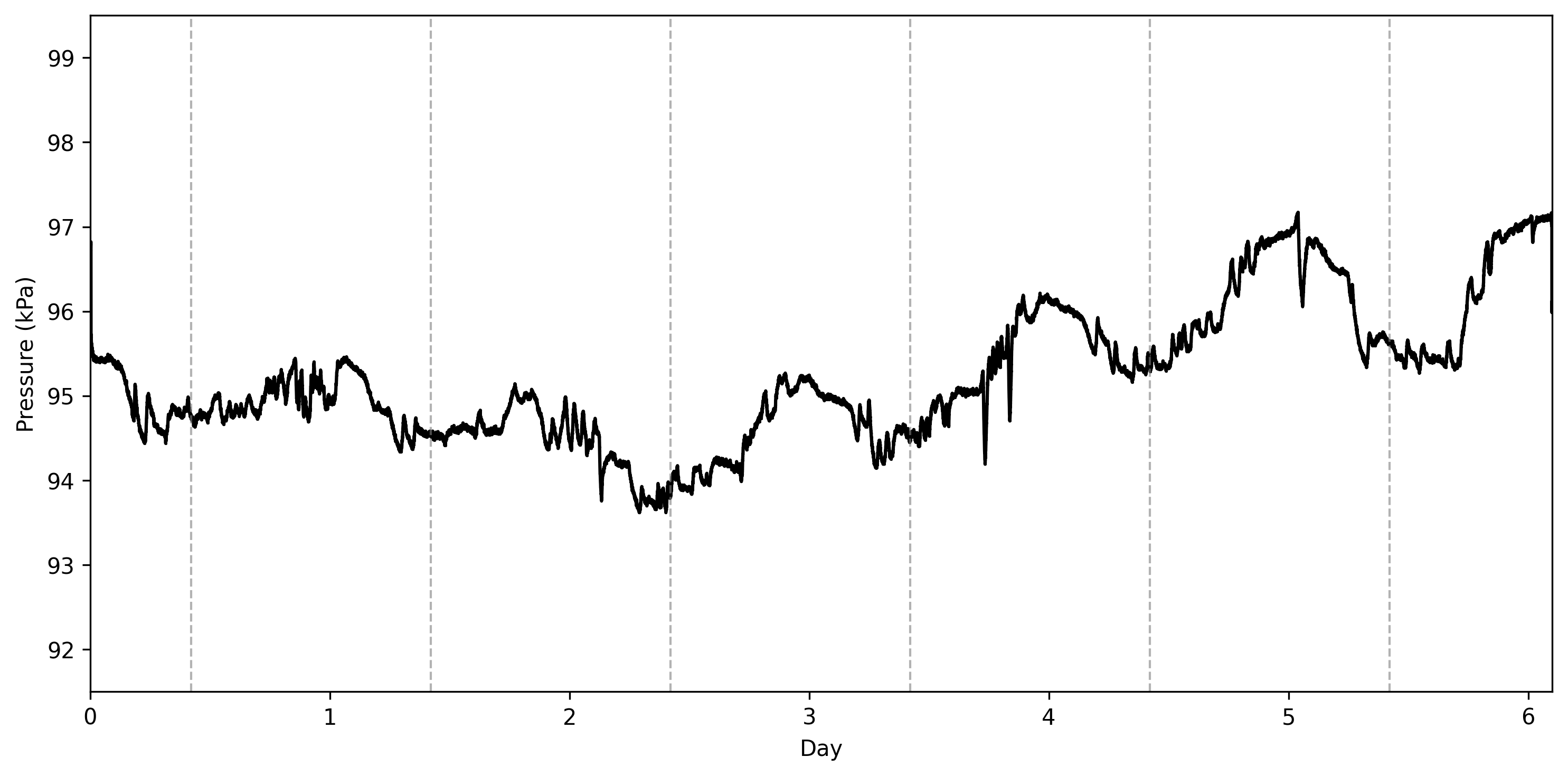}
\includegraphics[width=\columnwidth]{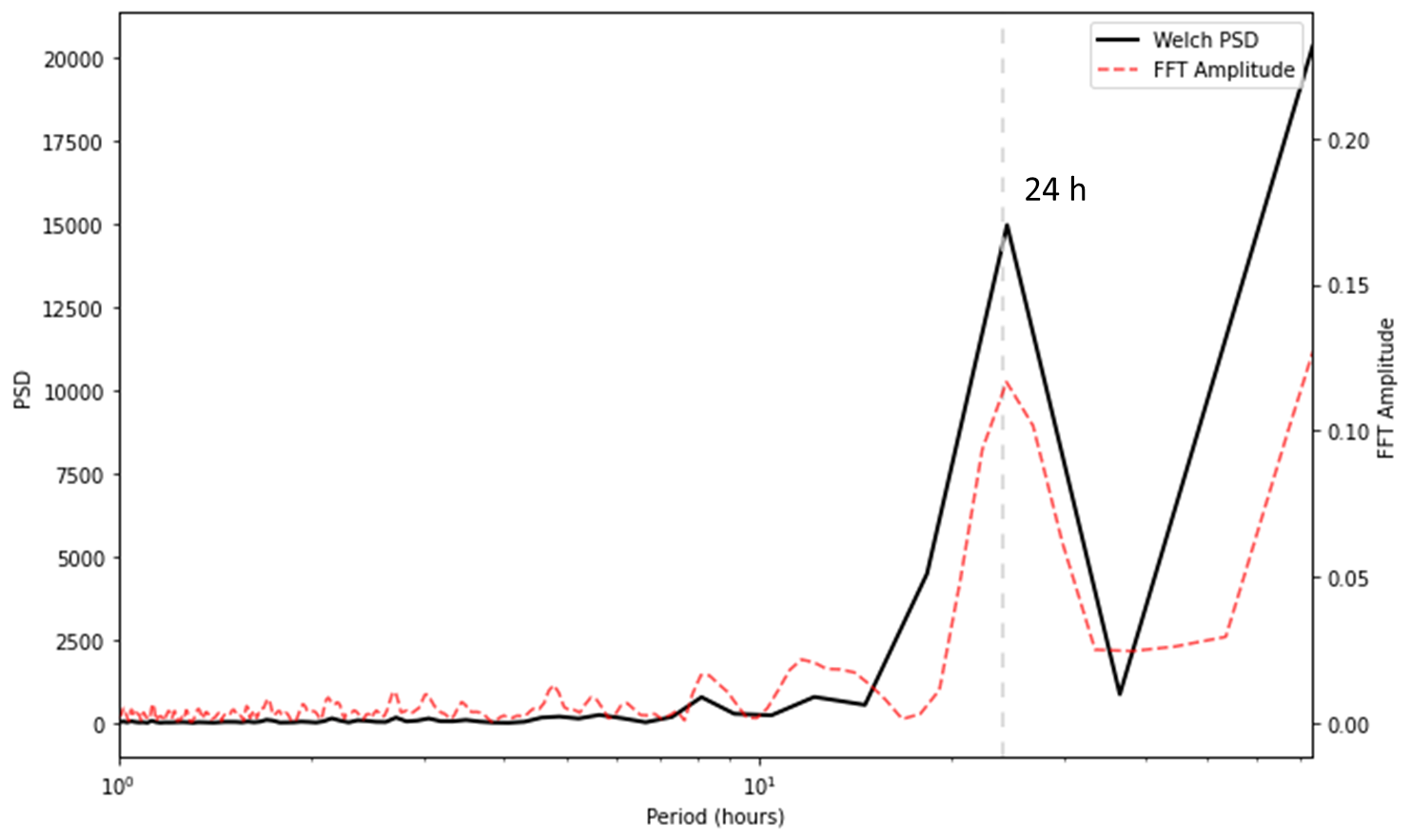}
\end{center}
Figure 3: (c) 
\end{figure}

\begin{figure}[tbp]
\begin{center}
\centering\includegraphics[width = \columnwidth]{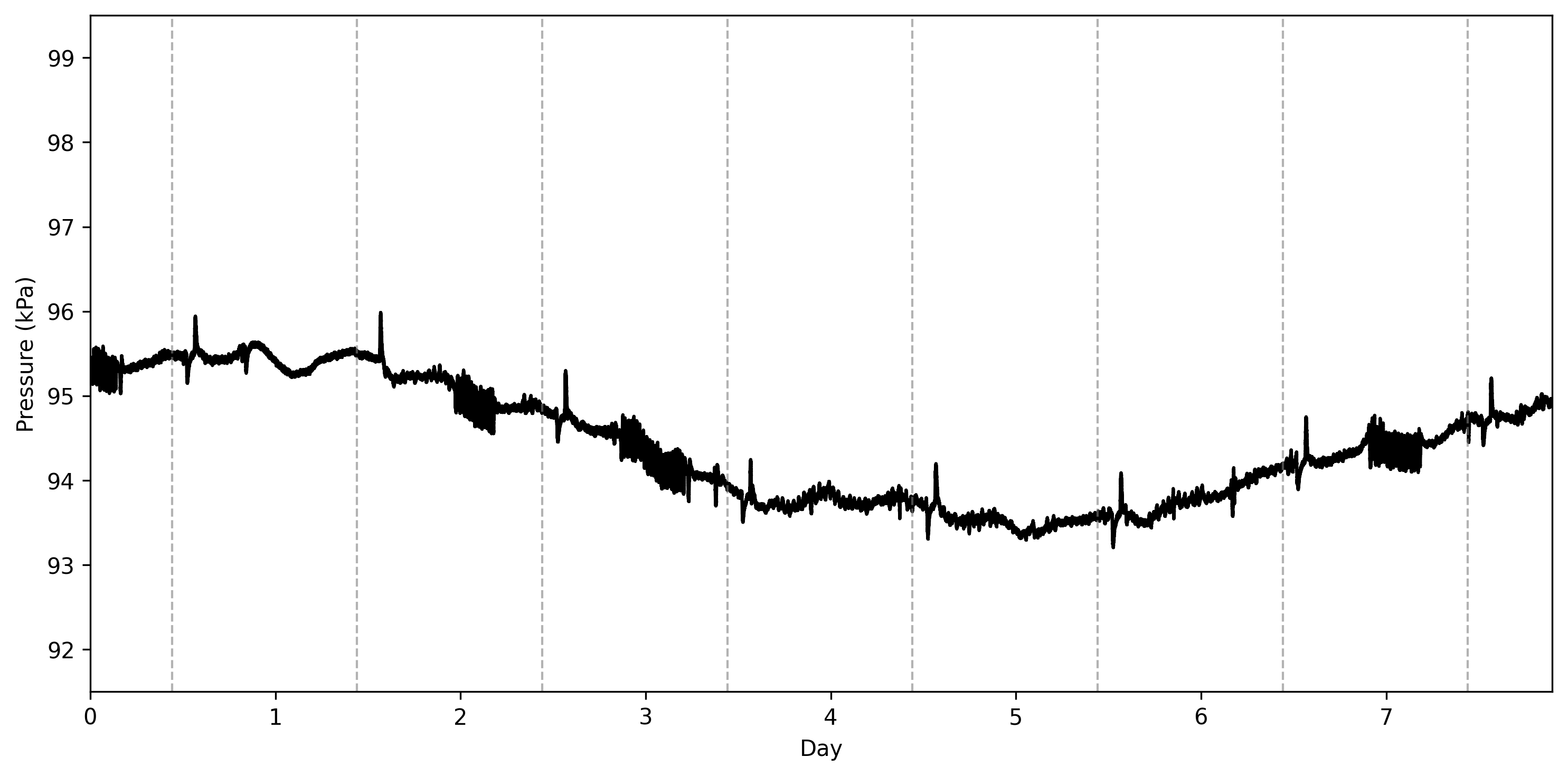}
\includegraphics[width=\columnwidth]{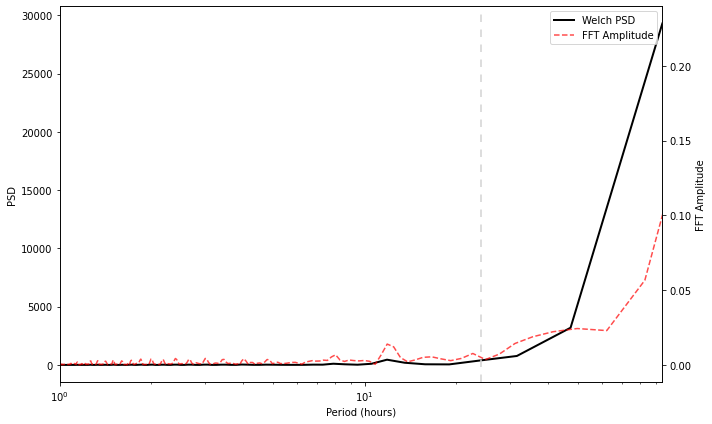}
\end{center}
Figure 3: (d)
\end{figure}

\begin{figure}[tbp]
\centering
\centering\includegraphics[width = \columnwidth]{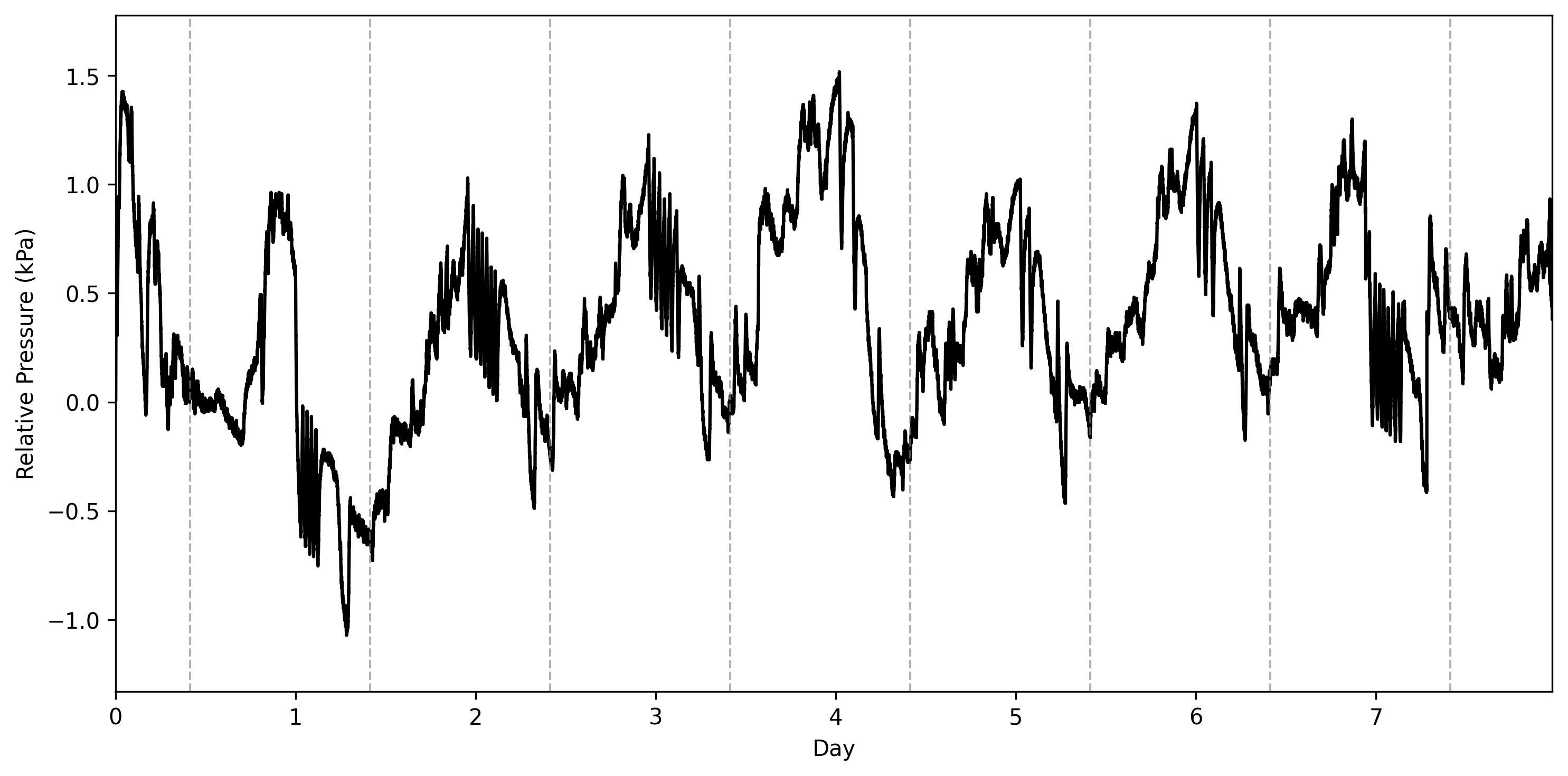}
\includegraphics[width=\columnwidth]{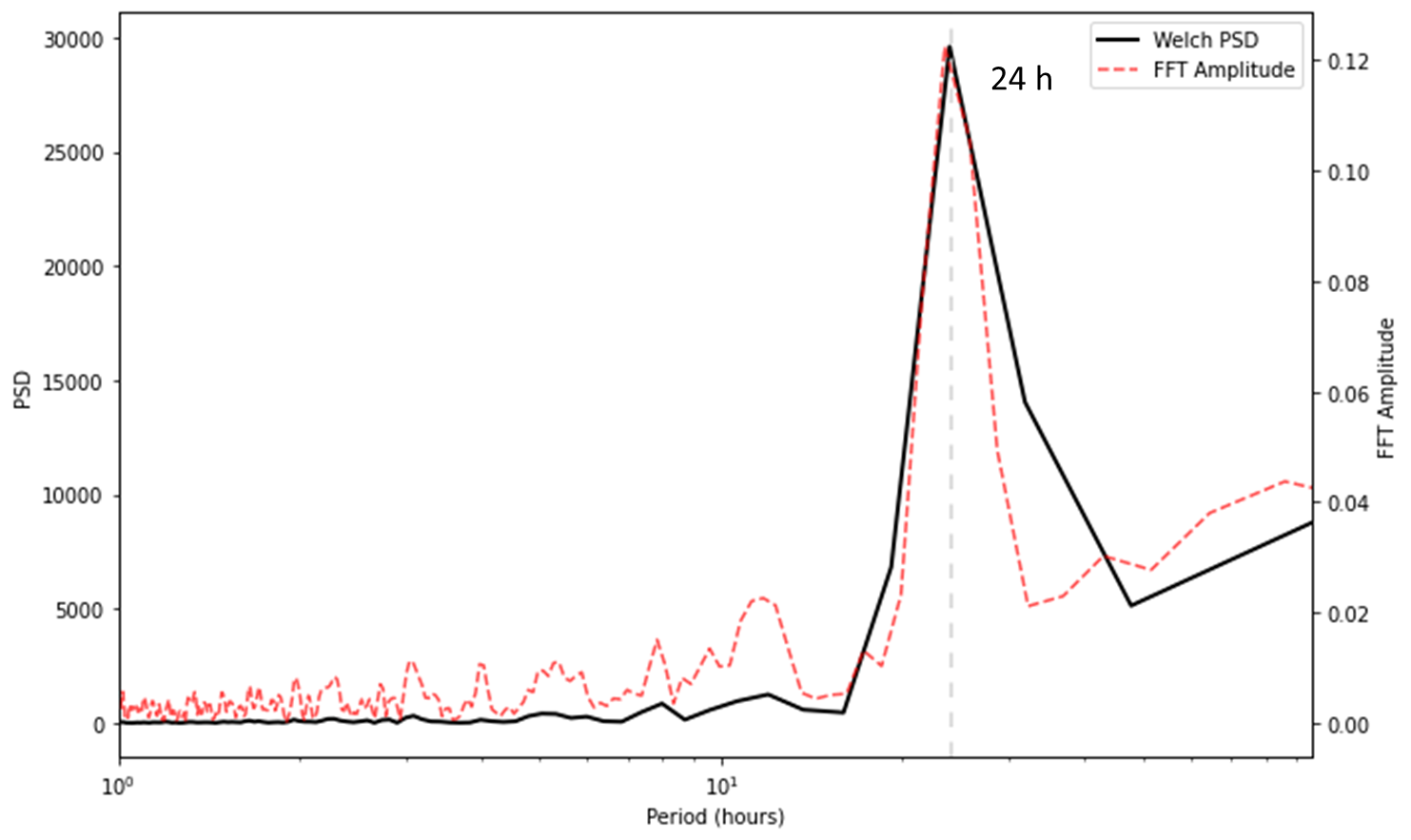}
\caption{(e) (a--e) (upper panels) Pressure time series inside \emph{Ecballium elaterium} fruits during their growth. 
Absolute pressure graphs (a--d) have been set to the same vertical scale to enable comparison; (e) is relative pressure. Midnight is marked by the dashed lines.
(lower panels) Fourier and Welch spectra of  the time series.
The circadian (24h) peak is marked with a dashed line. 
}\label{fig:fft-combined}\label{fig:pressure2}
\end{figure}

\begin{figure}[tbp]
\centering\includegraphics[width=\columnwidth]{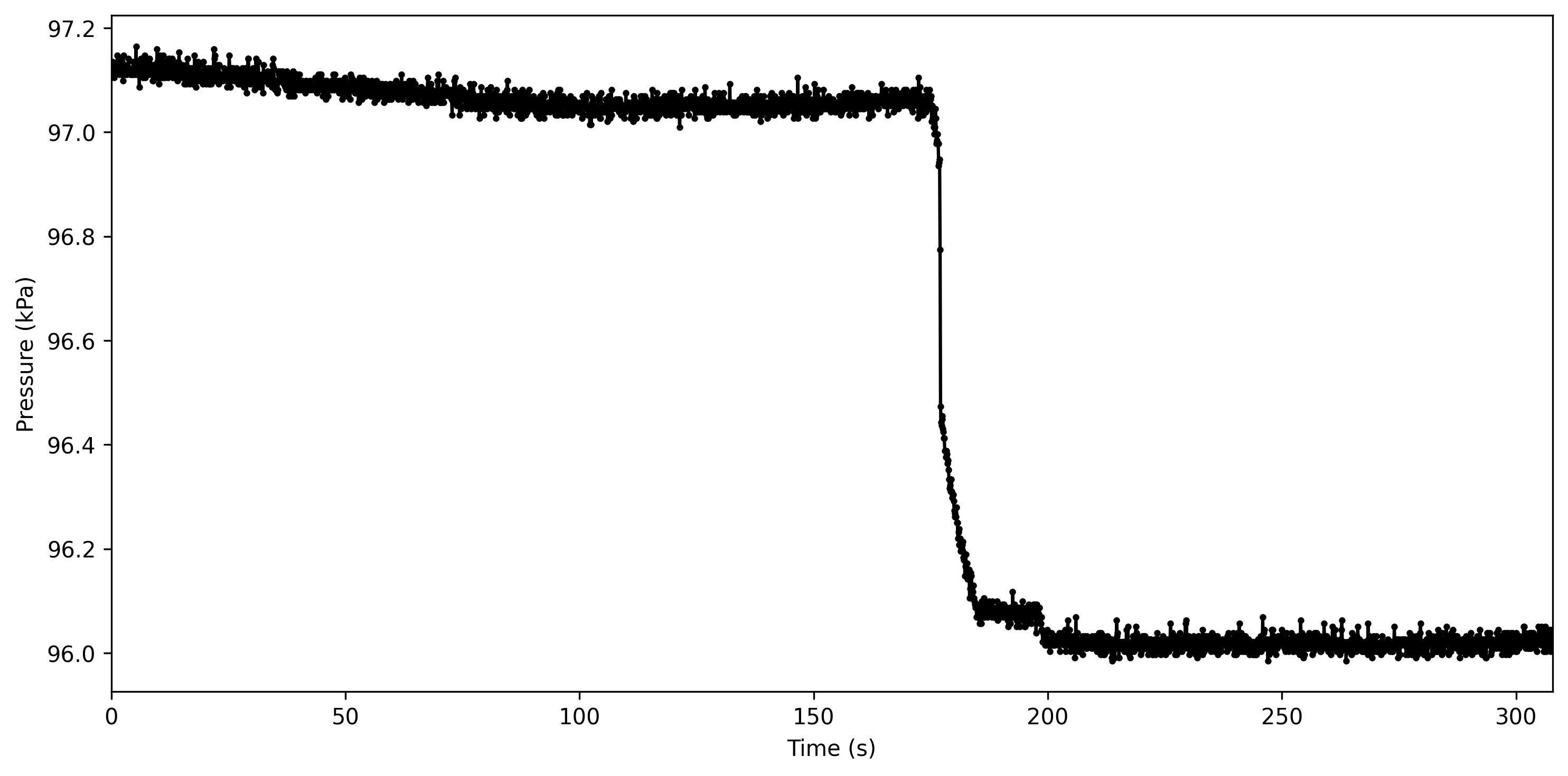}
\centering\includegraphics[width=\columnwidth]{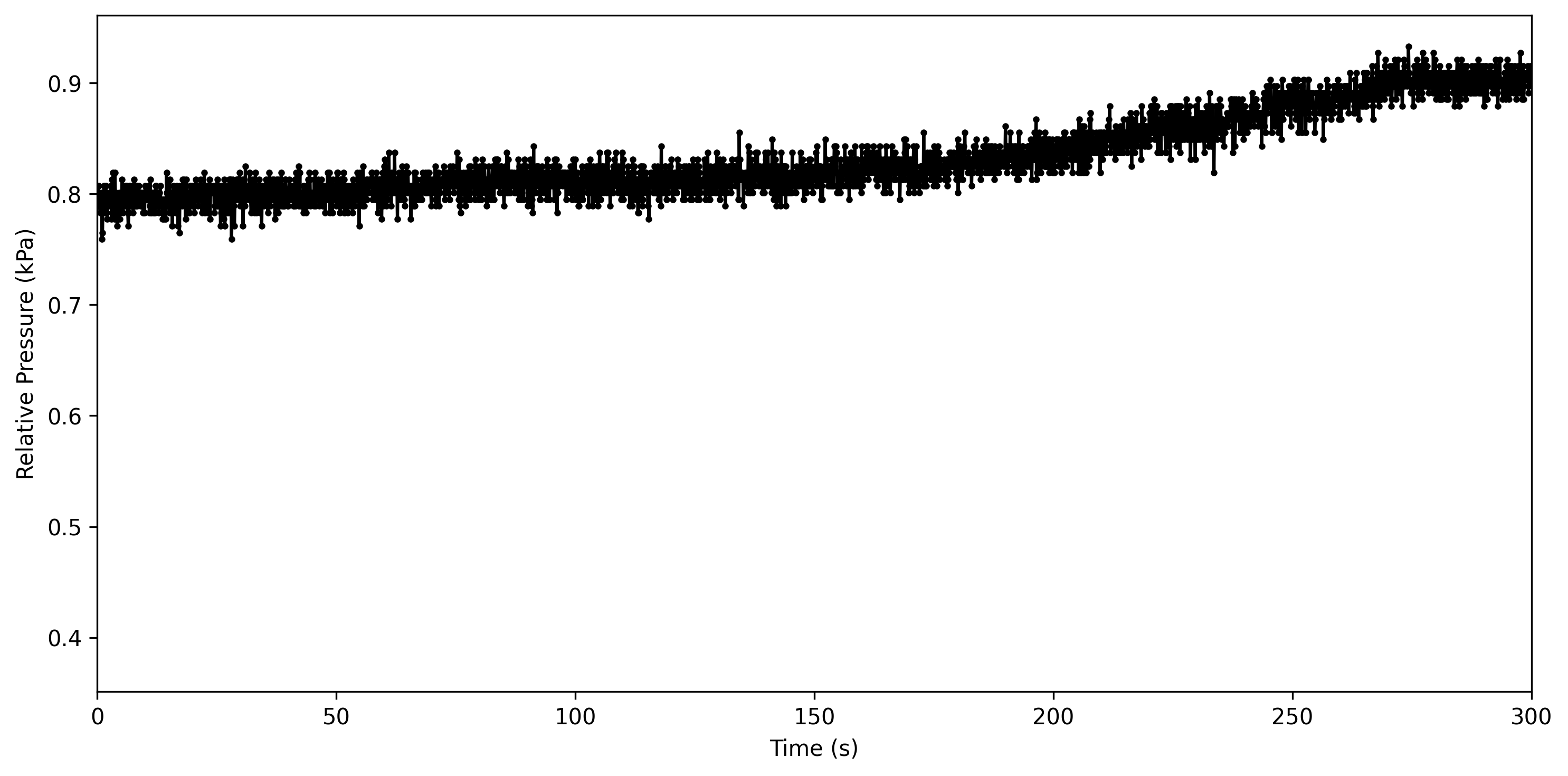}
\caption{Pressure measurements in the moment of explosion. a: Absolute pressure at the calyx end of the pod. b: Relative pressure measured at the same time in the stem above the fruit.
}\label{fig:explosion}
\end{figure}

We  recorded pressure time series  prior to and including explosions with  plants we took from the wild and  grew in the greenhouse. 
Figure~\ref{fig:pressure2}(a--d) shows  time series from different plants with the absolute pressure recorded via a sensor connected to the calyx (flower) end of the pod. The time series are taken over one to two weeks, all plotted with the same vertical axis to aid comparison. Figure~\ref{fig:pressure2}(e) shows
the same plant as in Figure~\ref{fig:pressure2}(d), recorded at the same time via a second sensor for relative pressure connected to the stem above the fruit. Lines are drawn at midnight, 00:00:00, for each graph. 
The first notable aspect of the time series is that they all appear very different, with very distinct waveforms. 

Let us consider first the diurnal cycle. 
In Fig.~\ref{fig:pressure2}(a)  the minimum of the diurnal pressure wave is close to midnight, always between 22:00 and 2:00.
However, in the case of Fig.~\ref{fig:pressure2}(b) the minima occur at 17:50 approximately,  with only minutes of variation. In Fig.~\ref{fig:pressure2}(c) and (e) the diurnal cycle is more variable than in (a), but the minima are  between 22:00 and 2:00. The  Fourier analyses of all these time series given in the lower panels show peaks at harmonics of one day, as expected. The Welch method corroborates. In the case of Fig.~\ref{fig:pressure2}(d) the diurnal cycle peak at 24h is missing in the Fourier representation; the 2nd harmonic at 12h is present but weak. The noise-reduced Welch method smooths out this feature completely, showing no diurnal periodicity. We can note in the signal of Fig.~\ref{fig:pressure2}(d) itself that the primary diurnal feature is a dip followed by a spike at the same time each morning. Although Fig.~\ref{fig:pressure2}(e) is recorded at the same time, this feature is not noticeable there. 

Now let us consider the tendency of each series. Figure~\ref{fig:pressure2}(a) and (c) show a rising tendency, but Figure~\ref{fig:pressure2}(b) and (e) are almost flat, and Figure~\ref{fig:pressure2}(d) is dominated by a weekly cycle. This long-period hebdomadal peak is seen in the Fourier and Welch analyses of (b), (c) and (d), but is hardly notable (a) and (e).
For the time series of Fig.~\ref{fig:pressure2}(a) and (d), there are small peaks in the Fourier spectra at 15 hours and its harmonics, but these are not present in the Welch representation, and they are most likely noise.
We see a great deal of small-scale structure in these pressure time series. Apart from the diurnal cycle, there is a large amount of short-period oscillations. Particularly noticeable are the short-period constant amplitude bursts lasting a couple of hours that are seen, for instace,  four times in Fig.~\ref{fig:fft-combined}(e), each time on the decay slope of the diurnal oscillation after its peak at midday, but only on some days. These bursts are also visible in Fig.~\ref{fig:fft-combined}(d).

From pressure data we can obtain a complementary picture of the explosion
to the video data we present below. Figure~\ref{fig:explosion} shows an example. This is a short five-minute time series, in which a ripe pod was exploded with the same technique as in the field, by a very light touch with a stick. Figure~\ref{fig:explosion}(a), which gives absolute pressure, is from the fruit and (b), which records relative pressure, from the stem.

\subsection{Video photography measurements}

\begin{figure}[tbp]
    \centering
    \includegraphics[width=0.31\textwidth]{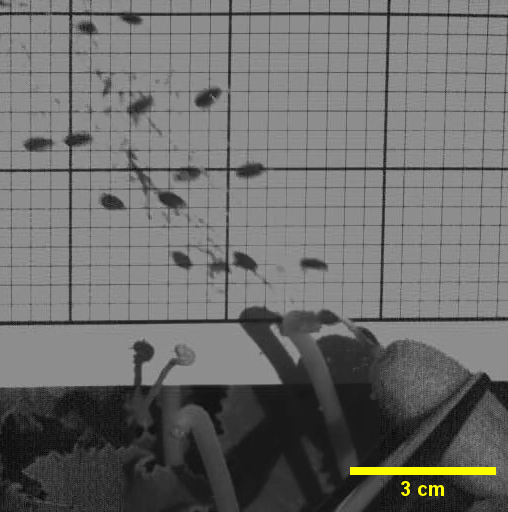}\hspace{0.02\textwidth}%
    \includegraphics[width=0.31\textwidth]{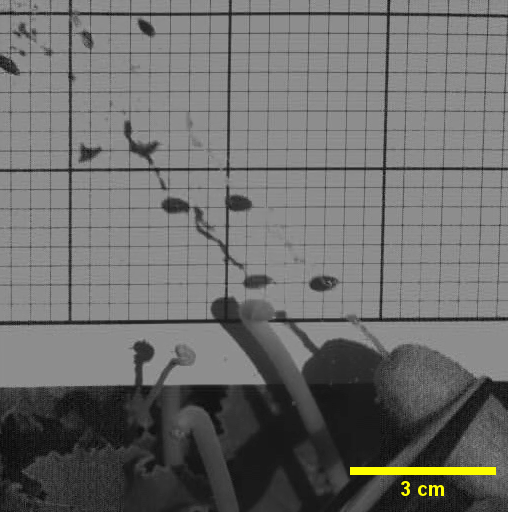}\hspace{0.02\textwidth}%
    \includegraphics[width=0.31\textwidth]{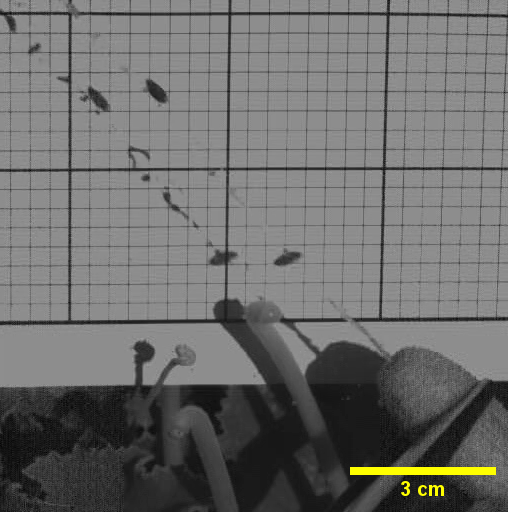} \\[0.3cm]
    \includegraphics[width=0.31\textwidth]{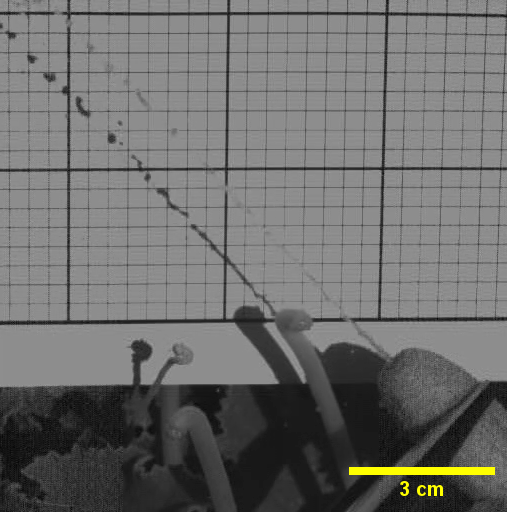}\hspace{0.02\textwidth}%
    \includegraphics[width=0.31\textwidth]{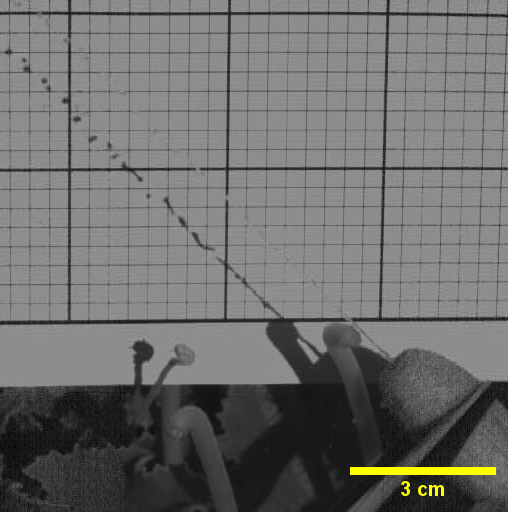}\hspace{0.02\textwidth}%
    \includegraphics[width=0.31\textwidth]{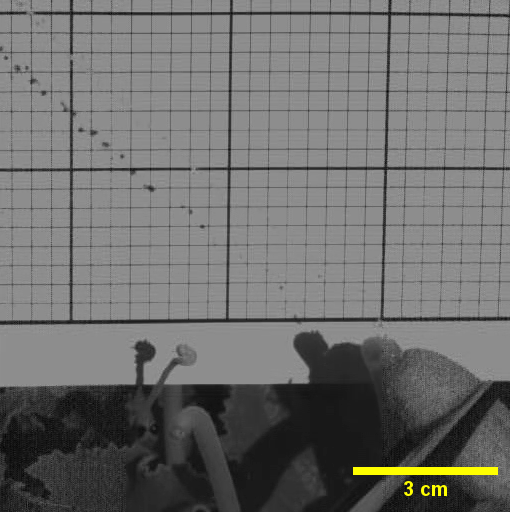} \\[0.3cm]
      \includegraphics[width=0.31\textwidth]{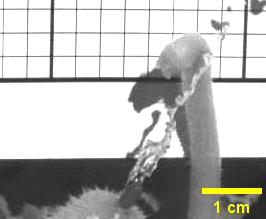}\hspace{0.02\textwidth}%
    \includegraphics[width=0.31\textwidth]{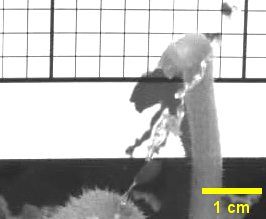}\hspace{0.02\textwidth}%
    \includegraphics[width=0.31\textwidth]{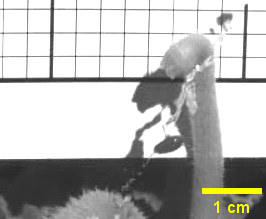} 
    \caption{(above) The explosion of an {\it Ecballium elaterium} seed pod seen at 1 ms intervals.
    (below) Closeup images demonstrating the seed orientation on exit.
}
    \label{fig:secuencia}
\end{figure}

\begin{figure}[tbp]
\centering\includegraphics[width=\columnwidth]{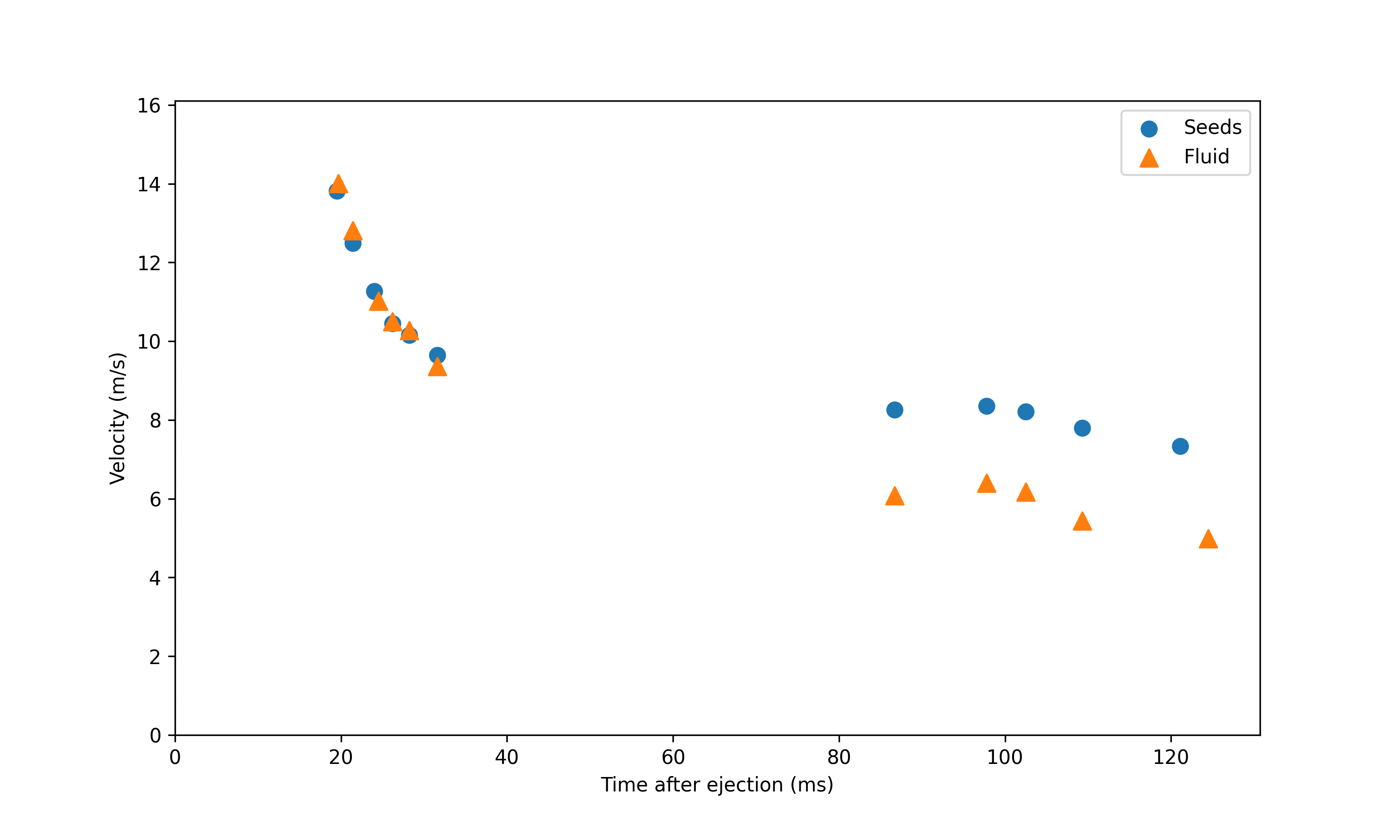}
\centering\includegraphics[width=0.86\columnwidth]{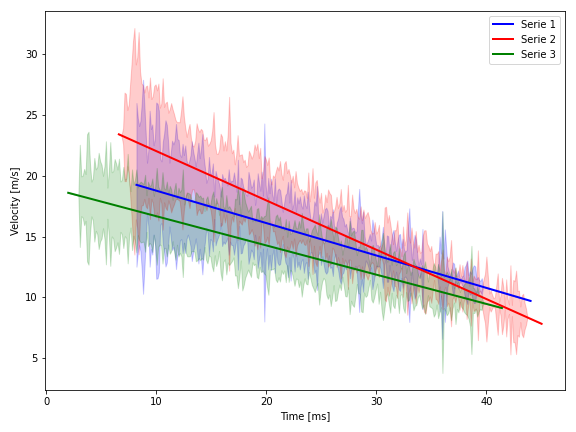}
\caption{
(a) Mean velocities of 11 different ejected seeds from the same plant, corresponding to series 3 in (b), ejected at times $t$ after the explosion and the fluid velocity at that time, using an ejected drop nearby. 
(b) Mean and standard deviation of seed velocities from 3 pods against exit time after the explosion.
}\label{fig:velocities}
\end{figure}

\begin{figure}[tbp]
\centering\includegraphics[width=\columnwidth]{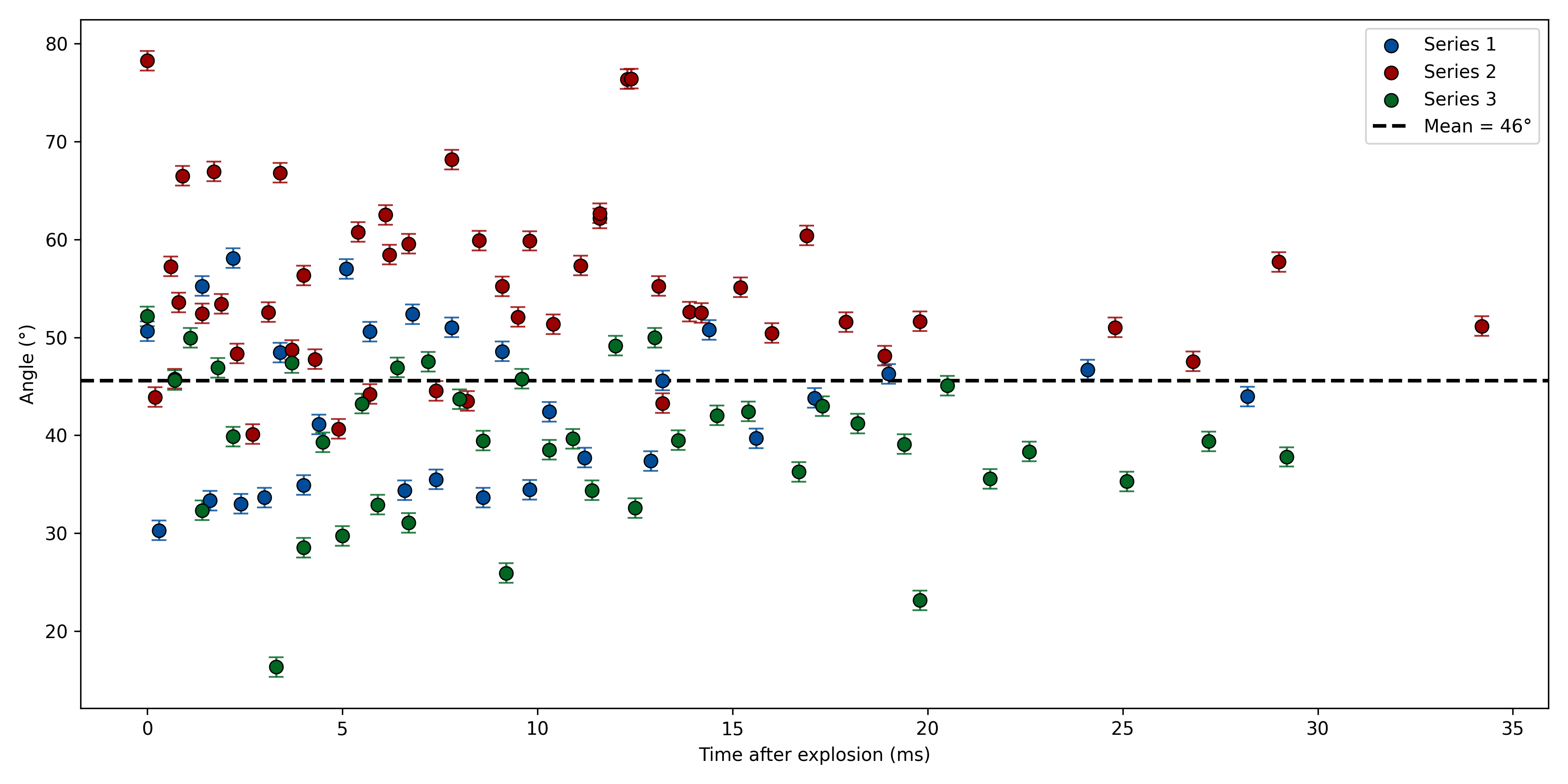}
\centering\includegraphics[width=\columnwidth]{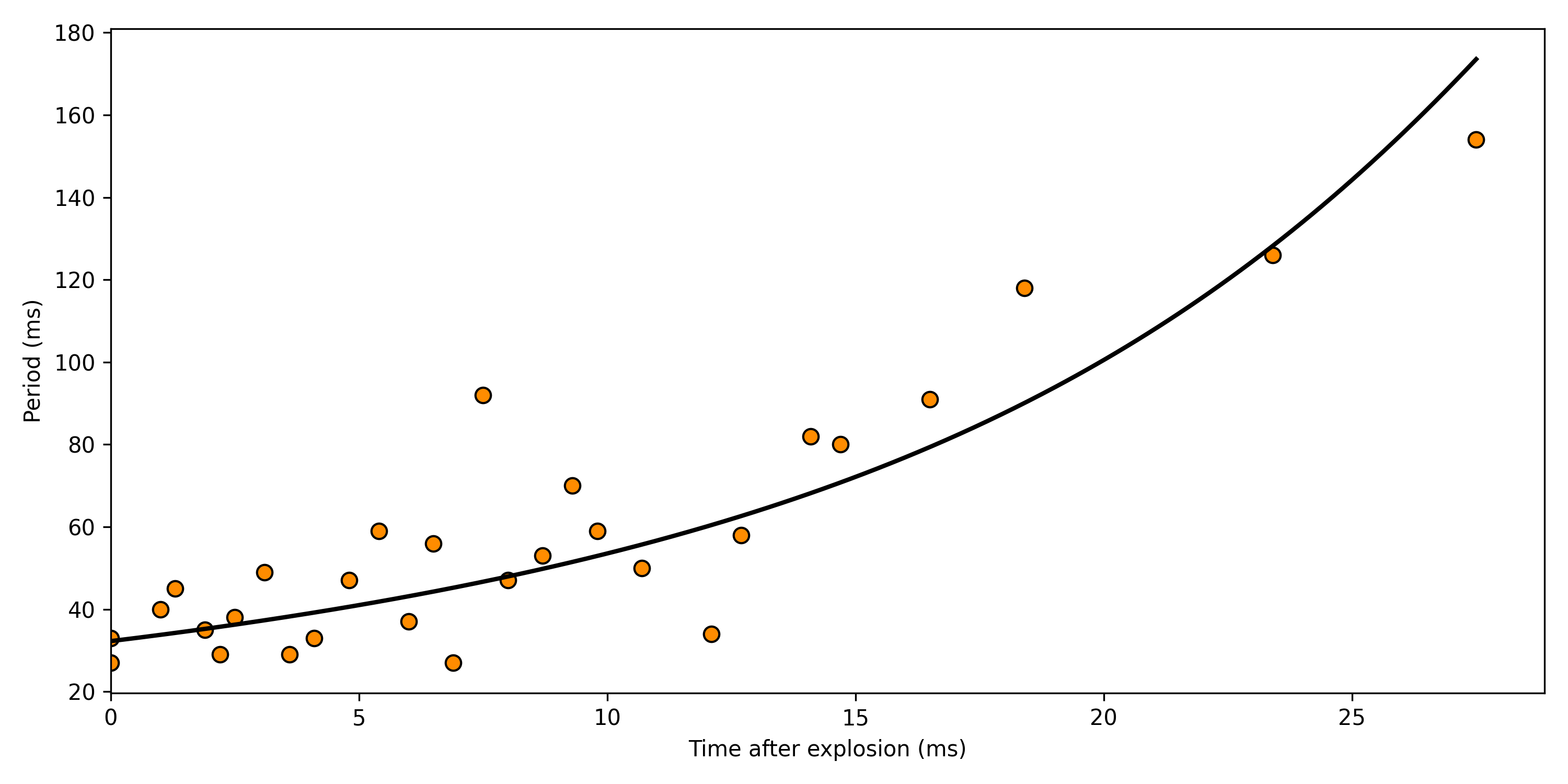}
\caption{(a) Seed ejection angles:  
angle from the horizontal  of
ejected seeds for three different seed pod explosions for seeds ejected at times $t$ after the explosion.
(b) Periods of  rotation of 30 different ejected seeds from the same plant ejected at times $t$ after the explosion. A curve has been added to guide the eye. 
}\label{fig:angles}\label{fig:rotation}
\end{figure}

We  recorded the explosion of  {\it Ecballium} seed  pods in the field  with a high-speed camera, as shown in Fig.~\ref{fig:field_trip}. 
From the videos---see  the supplementary data---we obtained data about both fluid and 
 seed trajectories. The pod breaks from the stem, leaving a circular orifice some 3--4~mm in diameter through which the seeds are ejected in a liquid jet. 
Note that no similar jet emerges from the broken stem section. 

The fluid flow in the  jet breaks up into a stream of drops, Fig.~\ref{fig:secuencia}. We plot in Fig.~\ref{fig:velocities}a the velocities of 9 different ejected drops
observed from 20 to 130 ms after the explosion. The drops are first ejected with a velocity of 14~m/s and over  120~ms of observation the ejection velocity slows to 6~m/s.
The fluid jet carries with it the seeds. 
At late times the seeds are faster than the fluid.
Note the two different slopes at early and late times, and that, for this pod, there was a intermediate period showing neither liquid nor seed ejection.

We counted the number of seeds ejected for 11 pod explosions; the minimum was 13 and the maximum 50, with a mean of 35 seeds per pod.
We  show in Fig.~\ref{fig:velocities}a the velocity data for the seeds. These data for three pod explosions show  discharge from 0--50~ms after the explosion, with the first seeds being expelled at up to 20~m/s, 19~m/s, and 17~m/s for the 3 pods, 
decaying to around 10~m/s for the  seeds expelled after 40~ms. Two of the pods discharged in a very similar way, having the same slope of velocity against time, while the third has a stronger discharge, with more speed at the start, but a shorter duration with a higher velocity/time slope.

 For our video photography we kept the pod fixed in place with a pair of tweezers. We should not forget that the seed pod, now separated from the plant, should also move under gravity and the reaction force from the jet. However, as the pod is much heavier than the seeds, its movement, even if it were  unrestrained, would be much slower than that of the jet. 

The seeds are ejected within a range of angles from the centre of the jet, like shot from a shotgun. 
Figure~\ref{fig:angles} indicates that initially 
 trajectories
gather around the edges of the dispersal cone, at angles up to 18\textdegree~  from the centre; later ejected seeds tend to have ejection angles towards the middle. 
We measured seed ejection angles from the horizontal for three pods, giving 16\textdegree--52\textdegree, 30\textdegree--58\textdegree, and 40\textdegree--78\textdegree.
When ejected, they exit the pod  aligned with the rounded (chalazal) end of their main axis first, as indicated in Fig.~\ref{fig:secuencia}.
The seeds also tumble. Figure~\ref{fig:rotation}b gives the rotation periods for 30 seeds from one pod explosion. The periods tend to increase for seeds ejected later in the explosion, from around 30 ms to 150 ms; a curve has been added to guide the eye.

\section{Discussion}

\subsection{Pressure time series}

A number of studies have looked at plant xylem pressures \cite{engelmann2004,thuermer1999,charrier2017}, but there is relatively little work looking in detail at such pressure time series. 
In our pressure measurements we find  
the pressures involved are between 92--99~kPa; almost one atmosphere. 
The explanation for the difference in the appearance of the  time series of Fig.~\ref{fig:pressure2} may lie in that they are taken in plants at different stages of seed development and ripening. 
We see the diurnal osmotic (turgor) pressure cycle in {\it Ecballium}, 
There is also a long-period variation that matches to a hebdomadal cycle. The explanation for this may be in human activities that occurred in weekly cycles in the greenhouse in which the plants were housed.
We had wished to capture the charging up to a maximum pressure before explosion. 
This charging tendency of increasing pressure over time is seen, but only in two of the four time series. Again, the explanation may lie in the plants being at different stages of ripening. 

Short-period cycles, or ultradian rhythms, in plants have been attested for many decades \cite{darwin1888power,bunning1956endogenous,sweeney2013rhythmic,mancuso2007rhythms}. Nonetheless, there is a dearth of knowledge regarding these oscillations, 
which must signal processes going on in the plant. Villalobos et al.~\cite{villalobos2025measuring} have noted in olive trees bursts with comparable period and duration to those we see in {\it Ecballium} in transpiration rate, xylem water potential, and root hydraulic conductance. We thus hypothesize that the short period processes we are seeing in 
\emph{E. elaterium} are of similar origin. 

We were able to capture the pressure change associated with the pod explosion, Fig.~\ref{fig:explosion}.
The pressure change seen in Fig.~\ref{fig:explosion}(a) is the absolute pressure recorded at the calyx, or flower end, of the seed pod. Before explosion, with an intact pod, this reads the almost 1 atmosphere pressure within the pod; after rupture it reads the external atmospheric pressure. The companion measurement in Fig.~\ref{fig:explosion}(b) is the relative pressure at the same time in the stem above the pod. Here no change is visible; we can surmise that the abscission mechanism \cite{jackson1972abscission}  prevents further sap loss after the pod breaks off from the stem. The pod pressure, Fig.~\ref{fig:explosion}(a), had fallen slightly to a minimum from which it had started to rise again before rupture. It is possible that eventually pressure rise alone might lead to rupture of a ripe pod, but in this instance---as must often be the case in the field---it was triggered by an external agent. On the other hand, the stem pressure measurement showed different, rising behaviour throughout. 

A notable aspect of both sets of simultaneous pod and stem pressure measurements,
Figs~\ref{fig:explosion}(a) and (b) and Figs~\ref{fig:pressure2}(d) and (e), is that the show little correlation. In the case of the week-long time series of Figs~\ref{fig:pressure2}(d) and (e), the fruit shows almost no circadian rhythm, but the stem shows a very strong circadian rhythm. The ultradian rhythms, on the other hand, are in some cases seen in both time series; in particular the four bursts noted in the afternoons of some of the days. Yet other processes, for instance the spikes each morning in Figs~\ref{fig:pressure2}(d), are not found in Figs~\ref{fig:pressure2}(e). This points to processes occurring only in one or other tissue of the plant.

\subsection{Fluid dynamics of the squirting}

The seed pod is in engineering terms a pressure vessel, as confirmed by the tomography study showing the large wall thickness; Fig.~\ref{fig:tomography}.
In engineering, liquids are often used to pressure test vessels such as boilers with a hydrostatic pressure test, since rupture by a liquid simply breaks the vessel wall, while a gas causes an explosion. The difference between a liquid and a gas explosion is because liquids are almost incompressible; that is to say a liquid has a fixed volume, while a gas expands to fix the volume available. Hence a gas explosion propels material outwards as the gas acts like a spring; a liquid explosion does not. What, then, is propelling the liquid jet here? The answer lies in the walls of the pressure vessel seen in the tomography: the spongy material of the wall stores energy as the pressure in the pod is increased. When the pod detaches, this material acting like a spring releases that energy, expelling the liquid and the seeds. Similar devices are used for energy storage in technological applications, where they are called hydraulic accumulators \cite{costa2023hydraulic}.

The action of seed expulsion is in fluid mechanical terms the pressure-driven flow of a particle-laden jet
from an orifice. This is moreover a three-phase flow, as there is the air surrounding the liquid jet, too; the overall dynamics is that of the breakup of a particle-laden liquid jet in air.
Using Bernoulli's principle gives a first estimate, ignoring viscous effects,  for  the jet velocity upon exit---sometimes called syringe flow---as   $U=\sqrt{2P/\rho}$, $P$ being the excess pressure and $\rho$ the liquid density. If we estimate one atmosphere overpressure, $P=10^5$~Pa, and $\rho=10^3$~kg/m$^3$, Bernoulli gives approximately 14~m/s.
Also from Bernoulli, the discharge rate $Q = C_d  A  \sqrt{2\Delta P / \rho}$ is, unlike the velocity, dependent on the orifice area $A$  (the prefactor $C_d$ adjusts for real-world losses). Thus we may interpret the results of Fig.~\ref{fig:velocities}(b). The initial velocity measured for the three pods, similar in two of the three, higher in the third, corresponds to  the initial pressure in the pods. The slope, again similar in two and higher in the third, shows a faster decrease in pressure $dP/dt$, which, given the fixed liquid volume in the pod, indicates a larger orifice in that case.

The jet is turbulent; we may compute its Reynolds number as $Re=U D/\nu$ where \(U\) is the characteristic velocity of the jet, \(D\) is the characteristic diameter of the jet, and \(\nu \) is the kinematic viscosity of the fluid. For $U\approx 14$~m/s, $D\approx 1$~cm and the viscosity of sap as 2--5 times water \cite{jensen2012modeling,jensen2013physical}, $\nu\approx2$--$5\times 10^{-6}$~m$^2$/s, we have $Re\approx 10^5$, which is well within the turbulent regime. 
To characterize the dynamics of this turbulent particle-laden liquid jet 
there are three variables we need to have in mind: the  particle size compared to the smallest turbulence scales, $D_p/\nu$; the particle-to-fluid density ratio, $\rho_p/\rho_f$; and the solid-phase volume fraction, $\Phi_V$ \cite{brandt2022particle}.
In our case, $D_p/\nu$ is large, $\rho_p/\rho_f$ is close to one, and $\Phi_V$ is also large. Of these three variables, it is the large particle size compared to the smallest turbulence scales, $D_p/\nu$, that puts us 
 into a ballistic dynamics regime, and one in which geometric effects---that is to say the seeds, by their size---affect the fluid flow. 
 
We see from Fig.~\ref{fig:secuencia} and from  the  videos  that the initial jet carrying the seeds breaks up into droplets rather rapidly, and the seeds become separated from the liquid. 
Breakup of a turbulent liquid jet is a complex process driven by the interplay between the liquid surface tension, viscosity, and the aerodynamic forces from the surrounding air \cite{phinney1973breakup,birouk2009liquid}.
In this case, there is also  the presence and distribution of the particles themselves. 
Compared to a fluid jet, the presence of particles alters the jet's initial deformation and later breakup stages, with higher effective viscosity slowing the process, while heterogeneity accelerates fragmentation of the jet by promoting rapid piercing of the liquid film. This breakup has been studied in detail for small particles \cite{Xu_Wang_Che_2023} but not, it seems, for large particles as we have here. The latter stages of the breakup involve a seed coated with a thin liquid coating that is stripped off under the flow through the air.
This breakup of the jet and separation of the seeds from the liquid droplets may account for the observation of  the seeds moving faster than the liquid drops at late times.

Ecballium has evolved a seed ejection mechanism that has not been used in human technology: by pressurizing and rupturing a pod containing the seeds with a liquid, the jet is incompressible.
Having the rupture form a nozzle, the relatively large particles that are the seeds are ejected ballistically within the liquid jet, and shed their liquid coating as they travel. This provides an efficient mechanism to expel all the seeds from the pod at as high a velocity as possible.

\subsection{Seed dispersal}

\begin{figure}[tbp]
\centering\includegraphics[width=0.8\columnwidth]{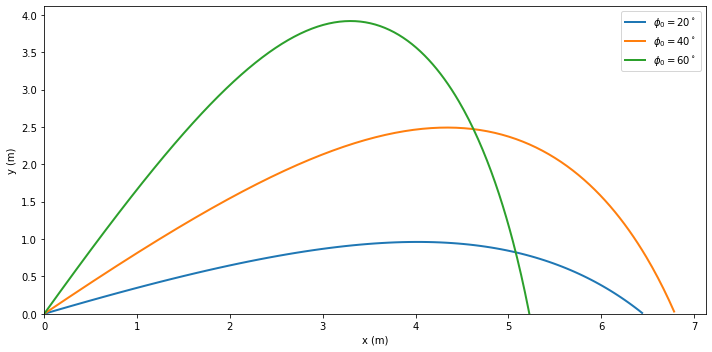}
\centering\includegraphics[width=0.8\columnwidth]{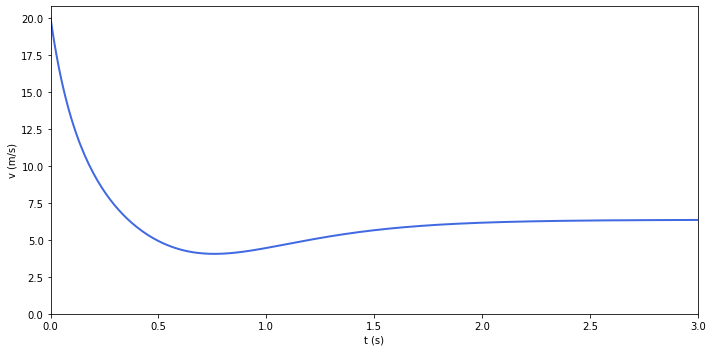}
\centering\includegraphics[width=0.8\columnwidth]{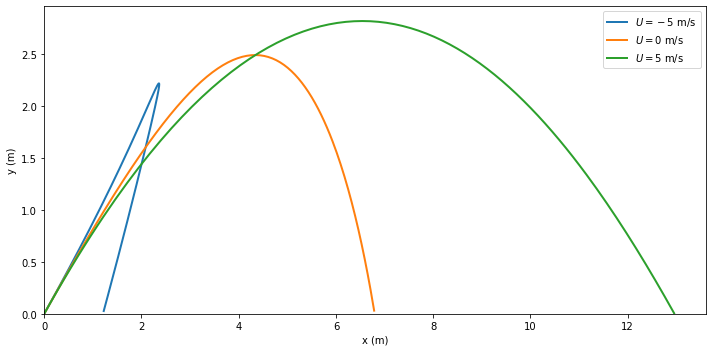}

\caption{Seed trajectories. 
(a) Trajectories for 3 different elevations for a starting velocity of 20~m/s; 
(b) velocity over time starting from 20~m/s tends rapidly to a terminal velocity in air;
(c) with no wind, a headwind of 5~m/s, and a tailwind of 5~m/s.
}\label{fig:trajectories}
\end{figure}

We triggered explosions using a light tap on the ripe pod. In nature, both similar touches by animals, and gusts of wind, would perform this function.
Also, we observed that if left undisturbed under greenhouse conditions, the maturing fruits reach a ripening stage in which abscission is endogenous, likely triggered hormonal mechanisms \cite{Li&Su2024-Abscission}. From  results on seed trajectories  and tomographic
scans of the pods  we can draw some conclusions about  the seed
dispersal  in \emph{E. elaterium}.
It seems that the arrangement of the seeds inside the pods is not
 random but helps the plant optimize seed detachment and scattering. Trajectories
gather throughout a dispersal cone (Fig.~\ref{fig:angles}a) and that
together with the stepped distribution inside the pod shown in Fig.~\ref{fig:tomography} leads to a spatial spread.

 For the ideal frictionless case, without taking drag into consideration, and with ejection at an elevation of 45\textdegree~from the horizontal, the range of a projectile is $V^2/g$; so for a seed travelling at 20 m/s, the maximum range in a vacuum would be 40~m. 
 However, this estimate neglects air drag.  When drag is added,  the angle for maximum range is somewhat less than 45\textdegree, and 
 a projectile under gravity and drag will tend towards a terminal velocity,  $v_t = \sqrt{2mg/({\rho C_d A)}}$, where $m$ is its mass, $g$ the acceleration due to gravity, $\rho$ the fluid density, and $C_d$ and $A$ the drag coefficient and frontal area of the projectile. For an {\it E. elaterium} seed,  this gives $v_t\approx 5$~m/s.
The  seed Reynolds number in air, $Re=U L/\nu$,
given that air has kinematic viscosity around $\nu=1.5\times10^{-5}$~m$^2$/s, that $L\approx 3$~mm, and $5\leq U\leq20$~m/s,  is approximately $10^3 \lesssim Re\lesssim 4\times 10^3$. Thus an {\it E. elaterium} seed falls  into the bottom end of the range of quadratic drag that acts for $Re> 10^3$  \cite{lubarda2022review}. 
 We plot graphs of some trajectories, under the approximation that the seeds are  spherical in shape,  in Fig.~\ref{fig:trajectories}.  Fig.~\ref{fig:trajectories}(a) shows the trajectories for three different angles of elevation for a seed ejected at 20~m/s.
Fig.~\ref{fig:trajectories}(b) demonstrates how the seed  tends rapidly to its terminal velocity.
Fig.~\ref{fig:trajectories}(c) shows that even a light breeze of 5~m/s (18~km/h) alters matters considerably; a headwind puts a seed almost back where it started, while a tailwind doubles the range. 

As seen in Figure~\ref{fig:angles}a, the mean angle of ejection is close to 45\textdegree, with a considerable spread. Angles of between 16\textdegree~and 78\textdegree~were noted, together with speeds between 10 and 30~m/s. From  
Fig.~\ref{fig:angles}a together with Fig.~\ref{fig:velocities}b, we see that the faster seeds at the start of the discharge had greater scatter in angle, while the slower ones towards the end were discharged with angles closer to the maximum range. 
The spread in exit velocities and angles is likely to provide effective bet-hedging \cite{childs_evolutionary_2010,starrfelt_bet-hedgingtriple_2012}. Seeds projected at low angles will hit against the maternal or neighbouring plant foliage, and will end up within the same maternal patch. Others, particularly those projected at higher angles and greater speeds, and with the benefit of wind, can travel tens of metres.  
Variation in dispersal distances will also entail a diversity of microenvironmental conditions, including shade, water, and nutrient availability. Ultimately, siblings of the same mother will experience uncorrelated abiotic and biotic (i.e., competition) environments, reducing the chance that all offspring fail simultaneously.

\section{Conclusions}

There are two aspects in which the {\it E. elaterium} seed dispersal mechanism is particularly notable notable:
1) The plant uses the physics of a particle-laden liquid jet, but one with very large particles; the liquid entrains the solid seeds, putting them into a ballistic regime. While technologies use particle laden liquid jets for abrasion and cutting, these are with much smaller particles, as well a much higher pressures, and moreover are, as discussed above, with a continuous supply. 
2) the plant organizes the explosion with just a small amount of liquid. Human technologies  use a liquid jet pumped with a continuous supply of liquid, as in a fire hose. But that would be extremely wasteful use of a resource for the plant, especially one in an arid environment. Instead it uses a form of hydraulic accumulator: a liquid pressurizes a pressure vessel with springy walls that store potential energy that when released, forms a liquid jet with only a small volume. 
In summary, the plant achieves the ejection of a small number of seeds at high velocity from a small pressure vessel using a minimal amount of liquid in a very efficient way. This sophisticated dispersal mechanism ensures bet-hedging by effectively spreading the progeny across space and environmental conditions. This constitutes a fascinating solution to area coverage that differs quite dramatically from human technologies.  

Although our work has provided further insight into the biomechanics of seed dispersal in {\it E. elaterium}, the causes and biological consequences of the osmotic pressure build up necessary remain hitherto unexplored. For instance, it has not been investigated whether the abiotic osmotic environment influences endogenous pressure. Is the osmotic build up homeostatic? Is it influenced by the osmolarity of the soil solution? Similarly, it is likely that  {\it E. elaterium} individuals differ in the osmotic potential of their fruit lumina. Is this variation plastic (i.e., influenced by external conditions)? What are its consequences for seed dispersal? Additionally, the fact that seeds develop inside a fruit with such high osmotic pressure might influence their physiology and ultimately the regeneration niche of the plant (i.e., the requirements for the replacement of one mature individual by a new mature individual of the next generation \cite{grubb1977}). For instance, are  {\it E. elaterium} seeds able to germinate under high osmotic potential or at least at higher than closely related species or species that grow in similar environments? These are all questions that illustrate the relevance of fluid dynamics in plant ecology.

\section*{Acknowledgements}

This work was made possible by grants nos.\ PID2024-160443NB-I00 and PID2022-143099OB-I00 financed by the Spanish Ministry and Agencia Estatal de Investigaci\'on MICIU/AEI/10.13039/501100011033 and European Regional Development Funds (ERDF), the European COST Action CA21169 Dynalife, supported by the EU Framework Programme Horizon 2020, the UGR intramural grant C-EXP-267-UGR23 and the European MSCA-SE-NANOTRIOSTEO Project 101182658.
    
\bibliographystyle{unsrt}
\bibliography{explosions}

@article{Bruhn2014,
  title={Osmosis-based pressure generation: dynamics and application},
  author={Bruhn, B. R. and Schroeder, T. B. H. and Li, S. and Billeh, Y. N. and Wang, K. W. and Mayer, M.},
  journal={PloS one},
  volume={9},
  pages={e91350},
  year={2014},
}

@article{deBruyn2015,
  title={Thermogenesis-triggered seed dispersal in dwarf mistletoe},
  author={deBruyn, R. A. J. and Paetkau, M. and Ross, K. A. and Godfrey, D. V. and Friedman, C. R.},
  journal={Nature Commun.},
  volume={6},
  pages={6262},
  year={2015},
}

@article{Dumais2012,
  title={``{Vegetable} dynamicks'': the role of water in plant movements},
  author={Dumais, J. and Forterre, Y.},
  journal={Annu. Rev. Fluid Mech.},
  volume={44},
  pages={453--478},
  year={2012},
}

@article{Forterre2013,
  title={Slow, fast and furious: understanding the physics of plant movements},
  author={Forterre, Y.},
  journal={J. Exper. Botan.},
  volume={64},
  pages={4745--4760},
  year={2013},
}

@article{Forterre2016,
  title={Physics of rapid movements in plants},
  author={Forterre, Y. and Marmottant, P. and Quilliet, C. and Noblin, X.},
  journal={Europhys. News},
  volume={47},
  pages={27--30},
  year={2016},
}

@article{Guo2015,
  title={Fast nastic motion of plants and bioinspired structures},
  author={Guo, Q. and Dai, E. and Han, X. and Xie, S. and Chao, E. and Chen, Z.},
  journal={J. Roy. Soc. Interface},
  volume={12},
  pages={20150598},
  year={2015},
}

@article{Hofhuis2016,
  title={Morphomechanical innovation drives explosive seed dispersal},
  author={Hofhuis, H. and Moulton, D. and Lessinnes, T. and Routier-Kierzkowska, A.-L. and Bomphrey, R. J. and Mosca, G. and Reinhardt, H. and Sarchet, P. and Gan, X. and Tsiantis, M. and Y. Ventikos and S. Walker and A. Goriely and R. Smith and A. Hay},
  journal={Cell},
  volume={166},
  pages={222--233},
  year={2016},
}

@article{Jensen2016,
  title={Sap flow and sugar transport in plants},
  author={Jensen, K. H. and  
  Berg-{Sorensen}, K. and Bruus, H. and Holbrook, N. M. and Liesche, J. and Schulz, A. and Zwieniecki, M. A. and Bohr, T.},
  journal={Rev. Mod. Phys.},
  volume={88},
  pages={035007},
  year={2016},
}

@article{Lewes1951,
  title={Observations on the internal pressure of the ripening fruit of \emph{Ecballium elaterium}},
  author={Lewes, D.},
  journal={Kew Bulletin},
  pages={443--444},
  year={1951},
}

@article{Obaton1947,
  title={Sur la projection des graines de l'\emph{Ecballium elaterium} {Rich}},
  author={Obaton, M. F.},
  journal={Bulletin de la Soci{\'e}t{\'e} Botanique de France},
  volume={94},
  pages={95--98},
  year={1947},
}

@article{Sakes2016,
  title={Shooting mechanisms in nature: a systematic review},
  author={Sakes, A. and van der Wiel, M. and Henselmans, P. W. J. and van Leeuwen, J. L. and Dodou, D. and Breedveld, P.},
  journal={PloS one},
  volume={11},
  pages={e0158277},
  year={2016},
}

@article{Skotheim2005,
author = "Skotheim, J. M. and  Mahadevan L.",
year =  2005,
title = "Physical limits and design principles for plant and fungal movements",
journal = "Science",
volume =  308, 
pages = "1308--1310",
}

@book{pliny,
author = "{Pliny the Elder}",
title = "Naturalis Historia (Natural History)",
publisher = "Book XX, 2",
year = "77--79 CE",
}

@article{janick2007,
title = "The Cucurbits of Mediterranean Antiquity: Identification of Taxa from Ancient Images and Descriptions",
author = "J. Janick and   H. S. Paris  and D. C. Parrish",
journal = "Ann. Botany", 
Volume = 100, 
year =  2007, 
Pages = "1441--1457",
}

@book{dioscorides,
author = "Dioscorides",
title = "De Materia Medica",
year = "50--70 CE",
}

@article{wolters1963,
author = "B. Wolters",
title = {{Hochfrequenzkinematiographische Untersuchung des Turgor-Sppritzmechanismus  von \emph{Ecballium}}},
journal = "Planta",
volume = 60, 
pages = "344--348", 
year = 1963,
}

@book{wolters_film,
author = "Wolters, B.",
title = "Ecballium elaterium (Cucurbitaceae) Ausschleudern der Samen",
publisher = "Film E 331, Inst. Wiss. Film. G{\"o}ttingen",
 year = 1963,
 }

@book{engelmann2004,
author = "Engelmann, W.",
title = "Rhythms in organisms --- Observing, experimenting, recording and analyzing",
year= 2004,
}

@article{thuermer1999,
author = "F. Th{\"u}rmer and J. J. Zhu and N. Gierlinger and H. Schneider and R. Benkert and P. Ge{\ss}ner and B. Herrmann and F.-W. Bentrup and U. Zimmerniann",
title = "Diurnal changes in xylem pressure and mesophyll cell turgor pressure of the liana \emph{Tetrastigma voinierianum}: The role of cell turgor in long-distance water transport",
journal = "Protoplasma",
volume = 206, 
pages = "152--162", 
year = 1999,
}

@article{charrier2017,
author = "Charrier, G. and Burlett, R. and Gambetta, G. A. and Delzon, S. and  Domec, J. C. and Beaujard, F.",
year =  2017,
title = "Monitoring Xylem Hydraulic Pressure in Woody Plants",
journal = "Bio-protocol",
volume = 7,
pages = "e2580",
}

@article{grubb1977,
  title={The maintenance of species-richness in plant communities: the importance of the regeneration niche},
  author={Grubb, P. J.},
  journal={Biol. Rev.},
  volume={52},
  number={1},
  pages={107--145},
  year={1977},
  publisher={Blackwell Publishing Ltd Oxford, UK}
}

@article{brandt2022particle,
  title={Particle-laden turbulence: progress and perspectives},
  author={Brandt, Luca and Coletti, Filippo},
  journal={Annual Review of Fluid Mechanics},
  volume={54},
  number={1},
  pages={159--189},
  year={2022},
  publisher={Annual Reviews}
}

@article{Xu_Wang_Che_2023, title={Breakup of particle-laden droplets in airflow}, volume={974}, DOI={10.1017/jfm.2023.847}, journal={Journal of Fluid Mechanics}, author={Xu, Zhikun and Wang, Tianyou and Che, Zhizhao}, year={2023}, pages={A42}}

@article{jensen2013physical,
  title={Physical limits to leaf size in tall trees},
  author={Jensen, Kaare H and Zwieniecki, Maciej A},
  journal={Physical review letters},
  volume={110},
  number={1},
  pages={018104},
  year={2013},
  publisher={APS}
}

@article{jensen2012modeling,
  title={Modeling the hydrodynamics of phloem sieve plates},
  author={Jensen, Kaare Hartvig and Mullendore, Daniel Leroy and Holbrook, Noel Michele and Bohr, Tomas and Knoblauch, Michael and Bruus, Henrik},
  journal={Frontiers in plant science},
  volume={3},
  pages={151},
  year={2012},
  publisher={Frontiers Research Foundation}
}

@article{costa2023hydraulic,
  title={Hydraulic accumulators in energy efficient circuits},
  author={Costa, Gustavo Koury and Sepehri, Nariman},
  journal={Frontiers in Mechanical Engineering},
  volume={9},
  pages={1163293},
  year={2023},
  publisher={Frontiers Media SA}
}

@article{bunning1956endogenous,
  title={Endogenous rhythms in plants},
  author={Bunning, E},
  journal={Annual Review of Plant Physiology},
  volume={7},
  number={1},
  pages={71--90},
  year={1956},
  publisher={Annual Reviews 4139 El Camino Way, PO Box 10139, Palo Alto, CA 94303-0139, USA}
}

@book{darwin1888power,
  title={The Power of movement in plants},
  author={Darwin, Charles and Darwin, Francis Ed},
  year={1888},
  publisher={John Murray}
}

@book{mancuso2007rhythms,
  title={Rhythms in plants: phenomenology, mechanisms, and adaptive significance},
  editor={Mancuso, Stefano and Shabala, Sergey},
edition="2nd",
  year={2015},
  publisher={Springer Science \& Business Media}
}

@book{sweeney2013rhythmic,
  title={Rhythmic phenomena in plants},
  author={Sweeney, Beatrice M},
  year={2013},
  publisher={Academic press}
}

@article{villalobos2025measuring,
  title={Measuring the Diurnal Variation of Root Conductance in Olive Trees Using Microtensiometers and Sap Flow Sensors},
  author={Villalobos, Francisco J and Testi, Luca and Garc{\'\i}a-Tejera, Omar and L{\'o}pez-Bernal, {\'A}lvaro and Tejado, In{\'e}s and Vinagre, Blas M},
  journal={Plant and Soil},
  volume={509},
  number={1},
  pages={999--1012},
  year={2025},
  publisher={Springer}
}

@article{jackson1972abscission,
  title={Abscission and dehiscence in the squirting cucumber, \emph{Ecballium elaterium}. {R}egulation by ethylene},
  author={Jackson, Michael B and Morrow, Ielene B and Osborne, Daphne J},
  journal={Canadian Journal of Botany},
  volume={50},
  number={7},
  pages={1465--1471},
  year={1972},
  publisher={NRC Research Press Ottawa, Canada}
}

@article{westermeier2018carnivorous,
  title={How the carnivorous waterwheel plant (\emph{Aldrovanda vesiculosa}) snaps},
  author={Westermeier, Anna S and Sachse, Renate and Poppinga, Simon and V{\"o}gele, Philipp and Adamec, Lubomir and Speck, Thomas and Bischoff, Manfred},
  journal={Proceedings of the Royal Society B: Biological Sciences},
  volume={285},
  number={1878},
  pages={20180012},
  year={2018},
  publisher={The Royal Society}
}

@article{guttenberg,
author = "Hermann von Guttenberg",
year = 1915,
title = "{Zur Kenntnis des Spritzmechanismus von \emph{Ecballium elaterium} Rich.}",
journal = "Berichte der Deutschen Botanischen Gesellschaft",
volume= 33,
pages = "20--37",
}

@article{overbeck1930druckkraften,
  title={{Mit welchen Druckkr{\"a}ften arbeitet der Schleudermechanismus der Spritzgurke? Untersuchungen an \emph{Ecballium elaterium} {R}ich}},
  author={Overbeck, Fritz},
  journal={Planta},
  volume={10},
  number={1},
  pages={138--169},
  year={1930},
  publisher={Springer}
}

@article{ziegenspeck,
author = "Hermann Ziegenspeck",
year = 1925,
title = "{{\"U}ber Zwischenprodukte des Aufbaues von Kohlenhydrat --- Zellw{\"a}nden und deren mechanische Eigenschaften}",
journal = "Botanisches Archiv. Zeitschrift für die gesamte Botanik",
volume = 9,
pages = "297--376",
}

@article{welch1967use,
  title={The use of fast {Fourier} transform for the estimation of power spectra: A method based on time averaging over short, modified periodograms},
  author={Welch, Peter D},
  journal={IEEE Transactions on audio and electroacoustics},
  volume={15},
  number={2},
  pages={70--73},
  year={1967}
}

@article{birouk2009liquid,
  title={Liquid jet breakup in quiescent atmosphere: A review},
  author={Birouk, Madjid and Lekic, Nebojsa},
  journal={Atomization and Sprays},
  volume={19},
  number={6},
  year={2009},
  publisher={Begel House Inc.}
}

@article{phinney1973breakup,
  title={The breakup of a turbulent liquid jet in a gaseous atmosphere},
  author={Phinney, Ralph E},
  journal={Journal of Fluid Mechanics},
  volume={60},
  number={4},
  pages={689--701},
  year={1973},
  publisher={Cambridge University Press}
}

@article{bowler2005causes,
  title={Causes and consequences of animal dispersal strategies: relating individual behaviour to spatial dynamics},
  author={Bowler, Diana E and Benton, Tim G},
  journal={Biological reviews},
  volume={80},
  number={2},
  pages={205--225},
  year={2005},
  publisher={Wiley Online Library}
}

@article{corlett2013will,
  title={Will plant movements keep up with climate change?},
  author={Corlett, Richard T and Westcott, David A},
  journal={Trends in ecology \& evolution},
  volume={28},
  number={8},
  pages={482--488},
  year={2013},
  publisher={Elsevier}
}

@book{cousens2008dispersal,
  title={Dispersal in plants: a population perspective},
  author={Cousens, Roger and Dytham, Calvin and Law, Richard},
  year={2008},
  publisher={Oxford University Press}
}

@article{hayashi2009mechanics,
  title={The mechanics of explosive seed dispersal in orange jewelweed (\emph{Impatiens capensis})},
  author={Hayashi, Marika and Feilich, Kara L and Ellerby, David J},
  journal={Journal of experimental botany},
  volume={60},
  number={7},
  pages={2045--2053},
  year={2009},
  publisher={Oxford University Press}
}

@article{schupp2019intrinsic,
  title={Intrinsic and extrinsic drivers of intraspecific variation in seed dispersal are diverse and pervasive},
  author={Schupp, Eugene W and Zwolak, Rafal and Jones, Landon R and Snell, Rebecca S and Beckman, Noelle G and Aslan, Clare and Cavazos, Brittany R and Effiom, Edu and Fricke, Evan C and Monta{\~n}o-Centellas, Flavia and others},
  journal={AoB Plants},
  volume={11},
  number={6},
  pages={plz067},
  year={2019},
  publisher={Oxford University Press US}
}

@article{travis2013dispersal,
  title={Dispersal and species' responses to climate change},
  author={Travis, Justin MJ and Delgado, Maria and Bocedi, Greta and Baguette, Michel and Barto{\'n}, Kamil and Bonte, Dries and Boulangeat, Isabelle and Hodgson, Jenny A and Kubisch, Alexander and Penteriani, Vincenzo and others},
  journal={Oikos},
  volume={122},
  number={11},
  pages={1532--1540},
  year={2013},
  publisher={Wiley Online Library}
}

@article{stamp1983ecological,
  title={Ecological correlates of explosive seed dispersal},
  author={Stamp, Nancy E and Lucas, Jeffrey R},
  journal={Oecologia},
  volume={59},
  number={2},
  pages={272--278},
  year={1983},
  publisher={Springer}
}

@article{nathan2000spatial,
  title={Spatial patterns of seed dispersal, their determinants and consequences for recruitment},
  author={Nathan, Ran and Muller-Landau, Helene C},
  journal={Trends in ecology \& evolution},
  volume={15},
  number={7},
  pages={278--285},
  year={2000},
  publisher={Elsevier}
}

@article{levin2003ecology,
  title={The ecology and evolution of seed dispersal: a theoretical perspective},
  author={Levin, Simon A and Muller-Landau, Helene C and Nathan, Ran and Chave,  J{\'e}r{\^o}me},
  journal={Annual Review of Ecology, Evolution, and Systematics},
  volume={34},
  number={1},
  pages={575--604},
  year={2003},
  publisher={Annual Reviews 4139 El Camino Way, PO Box 10139, Palo Alto, CA 94303-0139, USA}
}

@article{howe1982ecology,
  title={Ecology of seed dispersal},
  author={Howe, Henry F and Smallwood, Judith},
  journal={Annual review of ecology and systematics},
  volume={13},
  pages={201--228},
  year={1982},
  publisher={JSTOR}
}

@article{barea2022evolution,
  title={Evolution of fruit and seed traits during almond naturalization},
  author={Barea-M{\'a}rquez, Andr{\'e}s and Oca{\~n}a-Calahorro, Francisco J and Balaguer-Romano, Rodrigo and G{\'o}mez, Jos{\'e} Mar{\'\i}a and Schupp, Eugene W and S{\'a}nchez-P{\'e}rez, Raquel and Guillam{\'o}n, Jes{\'u}s and Zhang, Joanna and Rubio de Casas, Rafael},
  journal={Journal of Ecology},
  volume={110},
  number={3},
  pages={686--699},
  year={2022},
  publisher={Wiley Online Library}
}

@article{soomers2013wind,
  title={Wind and water dispersal of wetland plants across fragmented landscapes},
  author={Soomers, Hester and Karssenberg, Derek and Soons, Merel B and Verweij, Pita A and Verhoeven, Jos TA and Wassen, Martin J},
  journal={Ecosystems},
  volume={16},
  number={3},
  pages={434--451},
  year={2013},
  publisher={Springer}
}

@article{Li&Su2024-Abscission,
	author = {Li, Jiahuizi and Su, Shihao},
    Title = {Abscission in plants: from mechanism to applications},	
    Da = {2024/08/09},
	Doi = {10.1007/s44307-024-00033-9},
	Id = {Li2024},
	Isbn = {2948-2801},
	Journal = {Advanced Biotechnology},
	Number = {3},
	Pages = {27},
	Ty = {JOUR},
	Url = {https://doi.org/10.1007/s44307-024-00033-9},
	Volume = {2},
	Year = {2024}}

@article{starrfelt_bet-hedgingtriple_2012,
	title = {Bet-hedging---a triple trade-off between means, variances and correlations},
	volume = {87},
	copyright = {© 2012 The Authors. Biological Reviews © 2012 Cambridge Philosophical Society},
	issn = {1469-185X},
	url = {http://onlinelibrary.wiley.com/doi/10.1111/j.1469-185X.2012.00225.x/abstract},
	doi = {10.1111/j.1469-185X.2012.00225.x},
	language = {en},
	number = {3},
	urldate = {2014-10-24},
	journal = {Biological Reviews},
	author = {Starrfelt, Jostein and Kokko, Hanna},
	year = {2012},
	pages = {742--755}}

@article{childs_evolutionary_2010,
	title = {Evolutionary bet-hedging in the real world: empirical evidence and challenges revealed by plants},
	volume = {277},
	issn = {0962-8452},
	shorttitle = {Evolutionary bet-hedging in the real world: empirical evidence and challenges revealed by plants},
	url = {://WOS:000282645700001},
	doi = {10.1098/rspb.2010.0707},
	language = {English},
	number = {1697},
	journal = {Proceedings of the Royal Society B-Biological Sciences},
	author = {Childs, D. Z. and Metcalf, C. J. E. and Rees, M.},
	month = oct,
	year = {2010},
	pages = {3055--3064}}

@article{lubarda2022review,
  title={A review of the analysis of wind-influenced projectile motion in the presence of linear and nonlinear drag force},
  author={Lubarda, Marko V and Lubarda, Vlado A},
  journal={Archive of Applied Mechanics},
  volume={92},
  number={7},
  pages={1997--2017},
  year={2022},
  publisher={Springer}
}

\end{document}